\documentclass[11pt]{article}
\usepackage[letterpaper,margin=1in]{geometry}
\usepackage[numbers,sort&compress]{natbib}
\usepackage{lmodern}
\usepackage{enumitem}
\date{Preprint}
\usepackage[utf8]{inputenc} 
\usepackage[T1]{fontenc}    
\usepackage[hidelinks]{hyperref}       
\usepackage{url}            
\usepackage{longtable}
\usepackage{booktabs}       
\usepackage{amsfonts}       
\usepackage{nicefrac}       
\usepackage{microtype}      
\usepackage{xcolor}         
\usepackage{amsmath,amssymb,amsthm,mathtools}
\usepackage{subcaption}
\DeclareMathOperator*{\argmin}{arg\,min}
\newtheorem{lemma}{Lemma}
\newtheorem{assumption}{Assumption}

\newtheorem{theorem}{Theorem}

\newtheorem{corollary}{Corollary}
\usepackage{multirow}
\usepackage{array}
\usepackage{rotating}
\usepackage{pdflscape}
\usepackage{graphicx}
\usepackage{algorithm}
\usepackage{algpseudocode}
\title{Volatility-Corrected Prediction Intervals: Locally Adaptive Uncertainty Quantification for Probabilistic Time-Series Forecasting}
\author{Jason B. Cho \qquad David S. Matteson\\[0.5em]
\normalsize Department of Statistics and Data Science\\
\normalsize Cornell University, Ithaca, NY 14850\\
\normalsize\texttt{bc454@cornell.edu}\quad\texttt{dm484@cornell.edu}}

\begin{document}

\maketitle

\begin{abstract}
Prediction intervals from probabilistic time-series forecasters are commonly assessed by marginal coverage, the long-run proportion of observations falling inside the intervals. However, forecast errors in time series are often serially dependent and heteroskedastic, so prediction intervals can attain near-nominal marginal coverage while miscovering systematically during periods of larger local error variance. We identify and address this failure mode, which we call \emph{local scale miscalibration}. We first document the phenomenon using time-series foundation models (TSFMs): across several real-world series, their native intervals often attain near-nominal marginal coverage while exhibiting substantial temporal variation in local coverage, alternating between overcoverage and undercoverage over shorter windows. We then propose Volatility-Corrected Prediction Intervals (VCPI), a post-hoc calibration procedure that uses the base forecaster's predictive scale as a baseline, models the volatility of standardized forecast residuals with a Generalized Autoregressive Conditional Heteroskedasticity (GARCH) model, and applies a multiplicative correction to produce locally adaptive prediction intervals. We justify VCPI both theoretically and empirically, comparing it against model-agnostic conformal counterparts. We provide a theoretical analysis of interval adaptation of the proposed method following abrupt changes in volatility. Empirically, we evaluate VCPI on Chronos, Moirai, and TimesFM across four real-world datasets and find that it produces more stable local coverage than conformal baselines while maintaining near-nominal marginal coverage and comparable interval width.
\end{abstract}

\section{Introduction}\label{sec:intro}
Reliable time-series forecasts require both accurate point predictions and useful estimates of predictive uncertainty. Prediction intervals provide such estimates by giving a range of plausible future values. For a nominal miscoverage level \(\alpha\), a prediction interval is designed to contain the future observation with probability \(1-\alpha\). Prediction intervals are often evaluated in terms of their marginal coverage, the probability that the interval contains the future value after averaging over the data-generating process. This target is central to distribution-free methods such as conformal prediction, which provide finite-sample marginal coverage guarantees~\citep{conformal_original,cp}, and has been extended to sequential settings through methods such as adaptive conformal inference~\citep{conformal2} and quantile tracking~\citep{conformal_pid}. Empirical evaluations of probabilistic forecasts are also commonly reported as averages over the full test set, including empirical coverage, interval score~\citep{intervalscore}, continuous ranked probability score (CRPS)~\citep{crps}, and pinball loss~\citep{pinball}. 

These aggregate criteria are useful summaries, but they can hide poor uncertainty estimates in specific time periods or regimes. This is especially relevant for time-series forecasting, where forecast errors often exhibit temporal structure and may become larger or smaller during specific periods or regimes. As a result, prediction intervals can achieve the intended marginal coverage while still producing unreliable uncertainty estimates at a local level. This distinction is especially important for modern zero-shot probabilistic forecasters. Time-Series Foundation Models (TSFMs) are large-scale architectures trained on broad and diverse collections of time-series data, designed to support zero-shot forecasting across a wide range of domains~\citep{whatisfm}. Prominent examples include Chronos~\citep{chronos2}, TimesFM~\citep{timesfm}, Lag-Llama~\citep{lllama}, and Moirai~\citep{moirai2}. These models have demonstrated strong predictive accuracy across a variety of forecasting tasks and are increasingly being evaluated in applications such as finance~\citep{finance2,finance1}, energy forecasting~\citep{energy2,energy1}, demand forecasting~\citep{demand}, and climate and Earth system modeling~\citep{earth}, where uncertainty estimates are often most consequential during periods of stress, peak demand, or rapid regime change. Because TSFMs often provide predictive quantiles or samples from predictive distributions, their outputs are increasingly used not only for point forecasting but also for uncertainty quantification. 

\begin{figure}[t]
    \centering
    \includegraphics[width=0.9\linewidth]{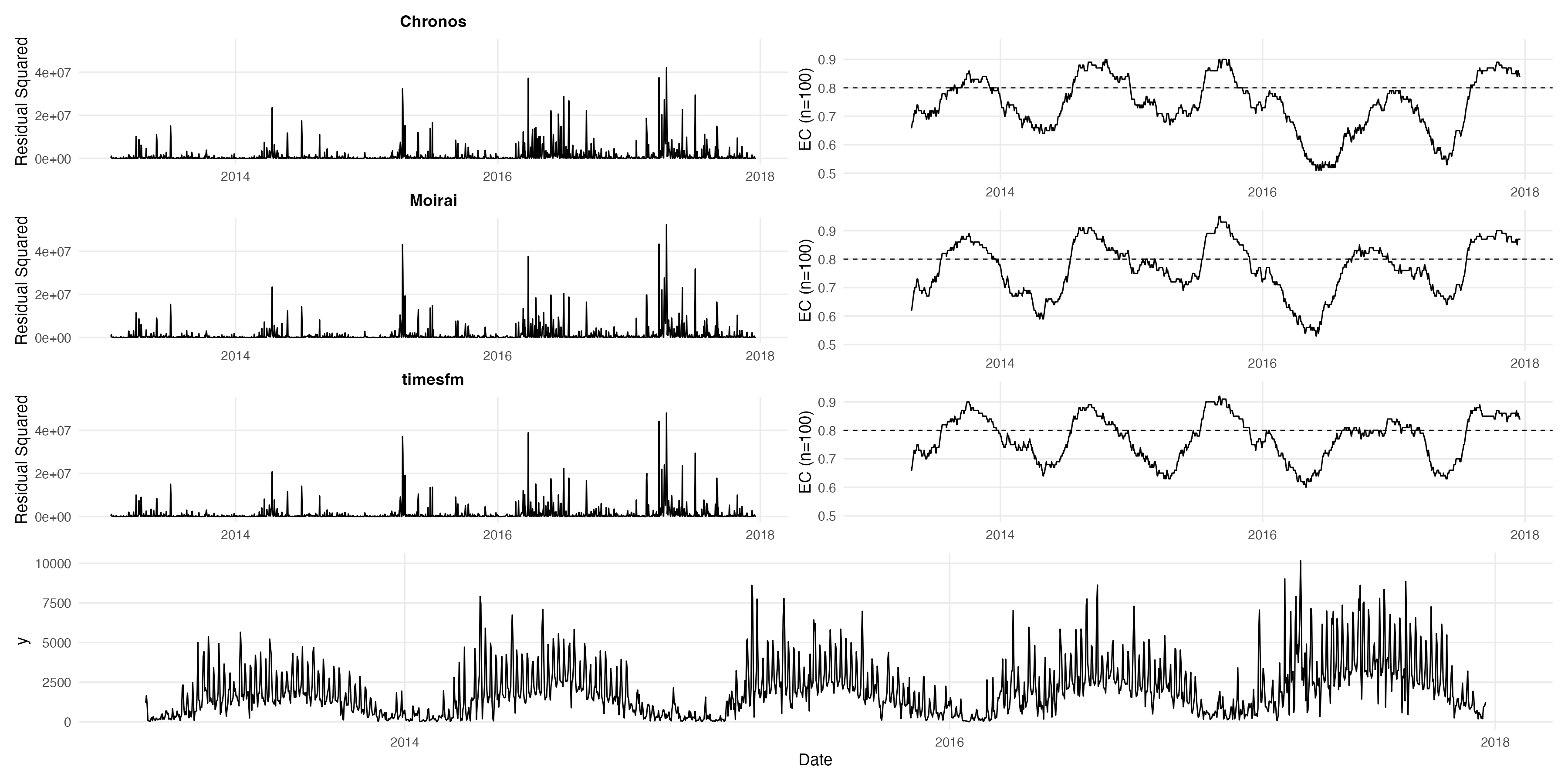}
	\caption{Daily bike rental counts from Capital Bikeshare~\citep{capital_bikeshare_data} (bottom), squared forecast residuals from three time-series foundation models: Chronos, Moirai 2, and TimesFM 2.5 (left column), and the empirical coverage of the models' prediction intervals, computed over 100 consecutive forecasts. The nominal 80\% coverage level is indicated by the horizontal dashed line (right column).}
    \label{fig:bike_residuals}
\end{figure}
To motivate our analysis, consider the bike rental data from Capital Bikeshare~\citep{capital_bikeshare_data}, together with rolling 16-step-ahead forecasts and their associated 80\% prediction intervals from three TSFMs, Chronos~\citep{chronos2}, Moirai 2~\citep{moirai2}, and TimesFM 2.5~\citep{timesfm} (Figure~\ref{fig:bike_residuals}). The observed series displays strong seasonal structure, with markedly higher prediction error during the summer months when demand peaks. The left panels of Figure~\ref{fig:bike_residuals} show that squared forecast residuals cluster during these same periods, indicating that forecast residuals increase systematically in high-demand regimes. In other words, forecast errors are not uniformly distributed over time but are conditionally larger when the volatility of the series itself is elevated. As a consequence, empirical coverage is also serially dependent. The right panels of Figure~\ref{fig:bike_residuals} show the empirical coverage of the nominal 80\% prediction intervals, which also exhibits a seasonal fluctuation and drops below 60\% in certain periods. 

In this paper, we propose the Volatility-Corrected Prediction Interval (VCPI) to address a specific failure mode that we call \emph{local scale miscalibration}, where marginal coverage may appear adequate even though interval reliability varies substantially across local error-scale regimes. VCPI uses the predictive scale produced by a base forecaster as a baseline, standardizes forecast residuals by this scale, fits a volatility model to the standardized residuals, and applies a multiplicative correction to future interval widths. We compare VCPI with model-agnostic uncertainty quantification methods for time-series forecasting, including adaptive conformal inference (ACI)~\citep{conformal2}, quantile tracking (QT)~\citep{conformal_pid}, Ensemble Batch Prediction Intervals (EnbPI)~\citep{enbpi}, and Sequential Predictive Conformal Inference (SPCI)~\citep{spci}. We also analyze the first-step GARCH update relative to QT after an upward volatility shift, conditional on a specified shock event. Empirically, we evaluate the methods across four real-world datasets using a rolling-window root mean squared coverage error (RMSCE), which measures deviations of local empirical coverage from the nominal coverage level. The results show that VCPI produces more stable local coverage than the conformal baselines while maintaining near-nominal marginal coverage.

\section{Related Work}\label{sec:related}
\paragraph{Volatility modeling for time series.} Modeling the conditional variance of a time series has a long history in econometrics. \citet{ARCH} introduced the ARCH model, in which the conditional variance is a function of past squared innovations, and \citet{GARCH2} generalized it to GARCH by also conditioning on past variances. A parallel strand of work models volatility as a latent stochastic process~\citep{Taylor,kimshephardchib1998,cho1}, leading to the stochastic volatility class of models. In applied forecasting, these specifications are typically combined with a conditional-mean model: ARMA-GARCH and its variants jointly model the mean and conditional variance of a series and are routinely used for interval forecasting in finance, energy, and climate applications~\citep{elec_armagarch, solar_armagarch}. Two features of this literature are central to our work. First, the conditional second moment, not just the conditional mean, is treated as a first-class forecasting target. Second, GARCH-type recursions react to new shocks immediately through the squared-residual term, a property that our theoretical analysis exploits in Section~\ref{sec:theory}. VCPI inherits this perspective but applies it to the \emph{forecast residuals} of a black-box probabilistic forecaster rather than to the data-generating process itself.

\paragraph{Conformal prediction for time series.} Conformal prediction~\citep{conformal_original,cp} provides distribution-free marginal coverage under exchangeability. Extending this framework to time series has motivated several lines of work: re-weighting approaches that accommodate non-exchangeability and distribution shift, including EnbPI~\citep{enbpi,tibshirani_covshift,barber2023conformal}; online methods that update a coverage-controlling parameter as observations arrive, including ACI~\citep{conformal2} and QT~\citep{conformal_pid}; methods designed for arbitrary distribution shifts~\citep{gibbs2024conformal}; and methods targeting local conditional coverage, such as SPCI~\citep{spci}. Despite their differences, these methods share a model-agnostic design philosophy that targets coverage by manipulating either the calibration set or a scalar coverage-control parameter. The temporal dynamics of the residual variance are not modeled explicitly.

\section{Volatility-Corrected Prediction Intervals (VCPI)}\label{sec:method}
In this section, we describe VCPI for one-step-ahead prediction: constructing a prediction interval for $Y_t$ using the information $\mathcal F_{t-1}$ available at time $t-1$. Appendix~\ref{app:vcpi_algorithm} provides the overall algorithm, including its extension to multi-step prediction.
\subsection{Reference Scale from Predictive Quantiles}\label{sec:ref}
Let $\hat f_t(\mathcal F_{t-1})$ denote the point forecast provided by a probabilistic forecaster for $Y_t$. Existing TSFMs characterize uncertainty around their forecast through either predictive draws~\citep{lllama} or predictive quantiles $\widehat Q_t(p\mid\mathcal F_{t-1})$, where $p\in(0,1)$ denotes the quantile level, such as deciles from the 10th to the 90th percentile~\citep{chronos2,moirai2,timesfm}.

VCPI first summarizes the uncertainty reported by the forecaster by extracting a single scale parameter, $\hat{\sigma}_{t}(\mathcal{F}_{t-1})$, from its predictive quantiles. Let $z_k=\Phi^{-1}(p_k)$, where $\Phi$ is the CDF of the standard normal distribution. Holding the supplied point forecast fixed, we fit a Gaussian reference scale to the reported quantiles:
\begin{equation}\label{eq:quant_optim}
\begin{aligned}
\hat{\sigma}_{t}(\mathcal F_{t-1}) &= \underset{\sigma\in\mathbb R}{\argmin}\; \sum_{k=1}^{K} \left[\widehat Q_t(p_k\mid\mathcal F_{t-1}) -\hat f_t(\mathcal F_{t-1}) -\sigma z_k \right]^2 \\ 
\end{aligned}
\end{equation}
$\hat{\sigma}_{t}(\mathcal F_{t-1})$ can be viewed as a weighted summary of paired interquantile spreads. It equals the standard deviation when the reported quantiles follow a Gaussian distribution. For non-Gaussian forecasts, it remains a summary of predictive spread. We focus on predictive quantiles because they are provided by the forecasters studied in our experiments. When predictive draws are available, the sample standard deviation may be used.

\subsection{Observation Model}
Consider first an idealized setting in which the conditional distribution of the observed data given the information available at forecast time $Y_t \mid \mathcal{F}_{t-1}$ is Gaussian and the forecaster correctly specifies its conditional mean and variance:
\[\mathbb E[Y_t\mid\mathcal F_{t-1}] = \hat f_t(\mathcal F_{t-1}), \qquad \operatorname{Var}(Y_t\mid\mathcal F_{t-1}) = \hat\sigma_t^2(\mathcal F_{t-1}).\]
Then the following model exactly represents this conditional distribution:
\begin{equation*}
Y_t = \hat f_t(\mathcal F_{t-1}) + \hat\sigma_t(\mathcal F_{t-1})\,\xi_t, \qquad \xi_t\mid\mathcal F_{t-1}\sim N(0,1).
\end{equation*}
In other words, dividing the forecast errors by $\hat\sigma_t(\mathcal F_{t-1})$ would yield standardized
errors with conditional variance one and a standard normal distribution. No further uncertainty correction would be needed.

As illustrated in Section~\ref{sec:intro}, however, the predictive quantiles reported by a forecaster may not fully capture uncertainty at a local level, and the standardized errors may not be Gaussian. VCPI introduces a multiplicative scale correction $s_t$ and allows a more general standardized error distribution $F$:
\begin{equation}\label{eq:vcpi_ob} Y_t = \hat f_t(\mathcal F_{t-1}) + \hat\sigma_t(\mathcal F_{t-1})\,s_t\,\xi_t,
\qquad \xi_t\mid\mathcal F_{t-1}\sim F.
\end{equation}
Here, $\hat f_t(\mathcal F_{t-1})$ is the forecast, and $\hat\sigma_t(\mathcal F_{t-1})$ summarizes the reported predictive spread, as described in Section~\ref{sec:ref}. The factor $s_t$ adjusts this scale over time, while $F$ is a mean-zero, unit-variance distribution describing the standardized error $\xi_t$ after this adjustment. This motivates two modeling tasks: forecasting the scale correction $s_t$ and estimating the error distribution $F$ from observed forecast errors.

\subsection{Modeling Residual-Scale Dynamics}
Based on the Equation~\ref{eq:vcpi_ob}, VCPI aims to model the normalized residual $e_{t}$, 
\begin{equation*}
e_t := \frac{Y_t-\hat f_t(\mathcal F_{t-1})} {\hat\sigma_t(\mathcal F_{t-1})} = s_t\xi_t
\end{equation*}
The sequence $\{e_t\}$ measures forecast errors relative to the uncertainty assumed by the forecaster. At time $t-1$, our goal is to forecast $s_t$ and estimate the standardized error distribution $F$ using the available information $\mathcal F_{t-1}$. This treats the variability of forecast errors as a forecasting target. 

Modeling errors through a time-varying scale and a standardized error distribution has a rich history in time-series analysis. ARCH and GARCH models describe conditional variance through past squared errors and past conditional variances~\citep{ARCH,GARCH2}. For example, the standard GARCH(1,1) specification is
\begin{equation}\label{eq:sGARCH}
s_t^2 = \omega+a e_{t-1}^2+b s_{t-1}^2,
\end{equation}
where $\omega>0$ and $a,b\geq0$ are parameters to be estimated. The squared-error term allows newly observed errors to affect the next variance forecast, while the lagged variance term captures persistence. When $a>0$, a larger absolute normalized error produces a larger subsequent scale correction, holding the previous variance fixed.

Extensions of GARCH provide flexibility in both the variance dynamics and the standardized error distribution. GJR-GARCH and EGARCH allow positive and negative errors of equal magnitude to have different effects on subsequent variance~\citep{gjrGARCH,egarch}. The distribution $F$ can likewise accommodate departures from normality. Student-$t$ distributions allow heavy tails~\citep{garch-t}, while skewed-$t$ distributions additionally allow asymmetry~\citep{garchfat}. Generalized error distributions provide another way to vary tail thickness relative to the normal distribution~\citep{egarch}.

These extensions make GARCH-family models a flexible framework for forecasting the scale correction $s_t$ and characterizing the standardized error distribution $F$. In implementation, the variance specification and error distribution $F$ are selected jointly using a temporally ordered validation segment within the calibration window. Candidate models are compared by validation coverage error. The selected specification is then refitted using the full calibration window to forecast $s_t$ and estimate $F$.

\subsection{Constructing the Corrected Intervals}
At time $t-1$, the fitted residual model provides a forecast of the scale correction, denoted by $\tilde s_t(\mathcal F_{t-1})$, and an estimated standardized error distribution, denoted by $\tilde F_{t-1}$. For a nominal miscoverage level $\alpha\in(0,1)$, VCPI constructs
\begin{equation}\label{eq:vcpi_interval}
\begin{split}
\widehat C_{t,\alpha}^{\mathrm{VCPI}}(\mathcal F_{t-1}) =\big[& \hat f_t(\mathcal F_{t-1}) +\hat\sigma_t(\mathcal F_{t-1}) \,\tilde s_t(\mathcal F_{t-1}) \,\tilde F_{t-1}^{-1}(\alpha/2),\\ & \hat f_t(\mathcal F_{t-1}) +\hat\sigma_t(\mathcal F_{t-1}) \,\tilde s_t(\mathcal F_{t-1}) \,\tilde F_{t-1}^{-1}(1-\alpha/2)\big].
\end{split}
\end{equation}
The interval combines the base forecast $\hat f_t(\mathcal F_{t-1})$, its reported spread summarized by $\hat\sigma_t(\mathcal F_{t-1})$, and the estimated residual model. Under the observation model in Equation~\eqref{eq:vcpi_ob}, if $s_t$ is determined by $\mathcal F_{t-1}$ and $F$ is continuous, using the true $s_t$ and $F$ in Equation~\eqref{eq:vcpi_interval} gives conditional coverage $1-\alpha$. The fitted procedure approximates this construction by forecasting the residual scale and estimating the standardized error distribution.

\section{Theory: Adaptivity After a Volatility Jump}\label{sec:theory}
VCPI makes a parametric modeling choice by using a GARCH model for the residual scale, whereas conformal methods provide model-agnostic marginal coverage guarantees. This section examines the benefit of this parametric assumption in a setting where local scale adaptation is necessary. Specifically, we consider a setting in which the scale of the forecast error increases abruptly. The relevant question is not only whether a method attains average coverage over time, but how quickly its interval width reacts to the new error scale. Under the assumptions below and conditional on the specified shock event, we show that, after a sufficiently large upward volatility jump, the GARCH-based VCPI update is closer to the post-jump oracle interval than the analyzed quantile tracking updates~\citep{conformal_pid} at the first prediction step after observing the jump. 
We retain the notation from Section~\ref{sec:method}; $e_t$ is the normalized forecast error, $s_t$ its scale, and $\xi_t$ the standardized error. For the theoretical analysis, consider the following data-generating process:
\[ e_t = s_t\xi_t, \qquad \xi_t\stackrel{\mathrm{iid}}{\sim}F,\]
where $s_t>0$ follows a deterministic baseline $v(t)$ and increases by $\Delta>0$ at time $\tau$:
\[
s_t= \begin{cases} 
	v(t), & t<\tau,\\
	v(t)+\Delta, & t\ge \tau.
\end{cases}
\]
We consider $\Delta\to\infty$, holding $\tau$, the calibration-window
length $W$, and the baseline path $v$ fixed. All methods construct
their next prediction interval at time $\tau$, after observing
$e_\tau$.

\begin{assumption}[Error Distribution]\label{as:error_dist}
$\{\xi_{t}\}_{t \geq 1}$ is assumed to be a mean zero, unit variance iid sequence of random variables with cumulative distribution function (CDF) $F$ that is symmetric around the mean. Let $G$ denote the CDF of $|\xi_{t}|$ with a density function $g$ that is continuous and strictly positive on $(0,\infty)$. $G^{-1}$ is used to denote the generalized inverse.
\end{assumption}
\begin{assumption}[Local Stability around $\tau$]\label{as:vt}
For some fixed $W\ge1$, there exists $\eta_{\tau,W}\ge0$ such that
\[
|v(t)-v(\tau)|\le\eta_{\tau,W},
\qquad t=\tau-W+1,\ldots,\tau+1.
\]
\end{assumption}
\begin{assumption}[Fitted GARCH Regularity]\label{as:garch}
The GARCH(1,1) coefficients $(\hat\omega,\hat a,\hat b)$, estimated using $\mathcal F_\tau$, satisfy the following bounds. There exist constants $0<\underline a,\underline b<1/2$, $0<\kappa<1$, and $0<\underline\omega<\infty$, independent of $\Delta$, such that
\[\underline a<\hat a<1-\underline a,\qquad \underline b<\hat b<1-\underline b,\qquad \hat a+\hat b\le1-\kappa,\qquad 0<\hat\omega\le\underline\omega.
\]
Let $\hat h_u$ denote the fitted variance state at time $u$, computed using the coefficients estimated at time $\tau$. The fitted variance recursion is initialized at
\[ \hat h_{\tau-W+1} =\frac{\hat\omega}{1-\hat a-\hat b}. \]
\end{assumption}

Fix a nominal miscoverage level $\alpha\in(0,1)$ and choose a calibration-window length $W>1/\alpha$. Let $c_\alpha:=G^{-1}(1-\alpha)$. The oracle half-width for the normalized forecast error $e_{\tau+1}$ is $q^\star_{\tau+1}:=c_\alpha s_{\tau+1}$. Using the scale-forecast notation from Section~\ref{sec:method}, define the GARCH half-width as
\[
\hat q^G_{\tau+1}(\mathcal F_\tau) =c_\alpha\tilde s_{\tau+1}(\mathcal F_\tau).
\]
The quantity $\hat q^{QT}_{\tau+1}(\mathcal F_\tau)$ denotes the corresponding half-width for $e_{\tau+1}$, constructed at time $\tau$ using quantile tracking~\citep{conformal_pid}.
\begin{lemma}[Asymptotic upper bound for GARCH(1,1)]
\label{lem:garch_asymptotic}
For any \(\rho\in(0,1)\), define
\begin{equation*}
E_{\tau}^{\mathrm{shock}}(\rho)
=\left\{\bigl|e_\tau^2-(v(\tau)+\Delta)^2\bigr|
\le
\rho (v(\tau)+\Delta)^2\right\}.	
\end{equation*}
Under the data-generating scheme in this section, Assumptions~\ref{as:error_dist}--\ref{as:garch} and conditional on \(E_{\tau}^{\mathrm{shock}}(\rho)\), as $\Delta\to\infty$,
\begin{equation*}
\bigl|\hat q_{\tau+1}^{G}(\mathcal F_\tau)-q_{\tau+1}^{\star}\bigr|
\le
c_\alpha\bigl[1-\underline a(1-\rho)\bigr]\Delta+O_p(1),
\end{equation*}
where \(\underline a>0\) is the deterministic lower bound on \(\hat a\) from Assumption~\ref{as:garch}. In particular, since \(0<\rho<1\) and \(\underline{a}>0\), the coefficient \(1-\underline{a}(1-\rho)\) is strictly smaller than \(1\).
\end{lemma}
\begin{proof}
	Appendix~\ref{app:lem1}
\end{proof}
As described in \citet{conformal_pid}, the QT update is defined as
\[
\hat q^{QT}_{\tau+1}(\mathcal F_\tau)
=
\hat q^{QT}_{\tau}(\mathcal F_{\tau-1})
+\gamma(\mathrm{miss}_\tau-\alpha),
\qquad
\mathrm{miss}_\tau
=
\mathbf 1\{|e_\tau|>\hat q^{QT}_{\tau}(\mathcal F_{\tau-1})\},
\]
where $\gamma>0$ is fixed independently of $\Delta$ and
$\hat q^{QT}_{\tau}(\mathcal F_{\tau-1})=O_p(1)$.
\begin{lemma}[Lower bound for quantile tracking]
\label{lem:aci_lower}
Under the data-generating scheme in this section, Assumptions~\ref{as:error_dist}--\ref{as:garch}, the following bound holds as $\Delta\to\infty$:
\begin{equation*}
\begin{aligned}
\bigl|\hat q_{\tau+1}^{QT}(\mathcal F_\tau)-q^\star_{\tau+1}\bigr|
&\ge c_\alpha\Delta-O_p(1).
\end{aligned}
\end{equation*}
Here, $O_p(1)$ denotes a nonnegative remainder bounded in probability, and $q^\star_{\tau+1}$ is the oracle half-width for $e_{\tau+1}$.
\end{lemma}
\begin{proof}
See Appendix~\ref{app:lem2}.
\end{proof}

\begin{theorem}[Asymptotic dominance]
\label{thm:asymptotic_dominance}
Under the data-generating scheme in this section, and Assumptions~\ref{as:error_dist}--\ref{as:garch} the following holds for any $\rho\in(0,1)$ and $\delta>0$:
\begin{align*}
&\mathbb P \bigg(
\bigl|\hat q_{\tau+1}^{QT}(\mathcal F_\tau)-q^\star_{\tau+1}\bigr|
-
\bigl|\hat q_{\tau+1}^{G}(\mathcal F_\tau)-q^\star_{\tau+1}\bigr|
\ge \delta
\,\bigg|\,
E_{\tau}^{\mathrm{shock}}(\rho)
\bigg)\to 1,
\end{align*}
as \(\Delta\to\infty\).
\end{theorem}
\begin{proof}
	Appendix~\ref{app:theorem}
\end{proof}
\begin{corollary}[Dominance under a linear lower bound]
\label{corr:dominance}
Under the same conditions as Theorem~\ref{thm:asymptotic_dominance}, let $\hat q_{\tau+1}(\mathcal F_\tau)$ be an interval half-width satisfying 
\[
\bigl|\hat q_{\tau+1}(\mathcal F_\tau)-q^\star_{\tau+1}\bigr|
\ge c_\alpha\Delta-O_p(1)
\]
as $\Delta\to\infty$. Then, for any $\rho\in(0,1)$ and $\delta>0$,
\[
\mathbb P\!\left(
\bigl|\hat q_{\tau+1}(\mathcal F_\tau)-q^\star_{\tau+1}\bigr|
-
\bigl|\hat q^G_{\tau+1}(\mathcal F_\tau)-q^\star_{\tau+1}\bigr|
\ge\delta
\,\middle|\,
E_\tau^{\mathrm{shock}}(\rho)
\right)\to1.
\]
\end{corollary}
\begin{proof}
Subtract the upper bound in Lemma~\ref{lem:garch_asymptotic}
from the assumed lower bound and apply the argument used in
Theorem~\ref{thm:asymptotic_dominance}.
\end{proof}

The theorem identifies a setting in which using error magnitude to forecast uncertainty gives VCPI an advantage. In the QT updates, a larger miss does not itself produce a larger update to their respective control parameters. It responds to whether the previous interval covered the observation. GARCH instead uses the squared error directly, allowing the magnitude of a newly observed shock to influence the next interval width.

\section{Empirical Results}
\begin{landscape}
\begin{table}[p]\caption{Reported metrics are marginal coverage (MC), rolling root mean squared coverage error (RMSCE), and median interval width across datasets and forecast horizon blocks. Columns compare the native model output, adaptive conformal inference (ACI), quantile tracking (QT), Ensemble Batch Prediction Interval (EnbPI), sequential predictive conformal inference (SPCI), and the proposed volatility-corrected prediction interval (VCPI), all targeting 80\% coverage (miscoverage level $\alpha = 0.2$). For ACI, the learning rate is fixed at 0.005, while for QT the learning rate multiplier is fixed at 0.05. RMSCE is the primary metric, as it measures local calibration under time-varying uncertainty. Blue bold entries indicate the smallest RMSCE within each model block. MC and Width are included  as complementary summaries and are not highlighted.}\label{tab:cov_80}
\centering\begingroup\fontsize{9}{11}\selectfont

\resizebox{\ifdim\width>\linewidth\linewidth\else\width\fi}{!}{
\begin{tabular}[t]{lllcccccccccccccccccc}
\toprule
\multicolumn{3}{c}{ } & \multicolumn{6}{c}{Chronos} & \multicolumn{6}{c}{Moirai} & \multicolumn{6}{c}{TimesFM} \\
\cmidrule(l{3pt}r{3pt}){4-9} \cmidrule(l{3pt}r{3pt}){10-15} \cmidrule(l{3pt}r{3pt}){16-21}
Data & Metric & Horizon & Native & ACI & QT & EnbPI & SPCI & VCPI & Native & ACI & QT & EnbPI & SPCI & VCPI & Native & ACI & QT & EnbPI & SPCI & VCPI\\
\midrule
 &  & 1 - 4 & 0.079 & 0.100 & 0.096 & 0.109 & 0.099 & \textbf{\textcolor{blue}{0.051}} & 0.083 & 0.105 & 0.102 & 0.111 & 0.104 & \textbf{\textcolor{blue}{0.051}} & 0.073 & 0.108 & 0.105 & 0.112 & 0.103 & \textbf{\textcolor{blue}{0.045}}\\

 &  & 5 - 8 & 0.082 & 0.092 & 0.087 & 0.099 & 0.095 & \textbf{\textcolor{blue}{0.069}} & 0.083 & 0.099 & 0.095 & 0.103 & 0.100 & \textbf{\textcolor{blue}{0.069}} & 0.075 & 0.101 & 0.099 & 0.107 & 0.100 & \textbf{\textcolor{blue}{0.059}}\\

 & \multirow{-3}{*}{\raggedright\arraybackslash RMSCE} & 9 - 16 & 0.092 & 0.106 & 0.099 & 0.112 & 0.109 & \textbf{\textcolor{blue}{0.081}} & 0.096 & 0.115 & 0.106 & 0.122 & 0.112 & \textbf{\textcolor{blue}{0.085}} & 0.080 & 0.111 & 0.107 & 0.118 & 0.105 & \textbf{\textcolor{blue}{0.077}}\\

 &  & 1 - 4 & 0.796 & 0.803 & 0.782 & 0.771 & 0.772 & 0.823 & 0.802 & 0.799 & 0.777 & 0.768 & 0.768 & 0.816 & 0.782 & 0.788 & 0.763 & 0.755 & 0.748 & 0.816\\

 &  & 5 - 8 & 0.804 & 0.803 & 0.782 & 0.762 & 0.753 & 0.839 & 0.810 & 0.799 & 0.776 & 0.750 & 0.747 & 0.832 & 0.792 & 0.788 & 0.766 & 0.747 & 0.738 & 0.817\\

 & \multirow{-3}{*}{\raggedright\arraybackslash MC} & 9 - 16 & 0.819 & 0.803 & 0.782 & 0.749 & 0.740 & 0.848 & 0.809 & 0.799 & 0.777 & 0.730 & 0.727 & 0.845 & 0.803 & 0.792 & 0.768 & 0.728 & 0.727 & 0.846\\

 &  & 1 - 4 & 1947.250 & 1929.987 & 1802.969 & 1796.551 & 1751.282 & 2059.074 & 1870.249 & 1831.740 & 1756.069 & 1756.724 & 1737.682 & 1860.122 & 2050.737 & 1952.828 & 1916.520 & 1910.394 & 1882.921 & 2261.180\\

 &  & 5 - 8 & 2280.500 & 2045.242 & 1858.352 & 1884.949 & 1812.761 & 2354.316 & 2114.440 & 1974.620 & 1846.830 & 1826.787 & 1797.731 & 2106.158 & 2290.387 & 2150.538 & 2035.227 & 2039.196 & 2012.662 & 2488.308\\

\multirow{-9}{*}{\raggedright\arraybackslash Bike Rental Counts} & \multirow{-3}{*}{\raggedright\arraybackslash Width} & 9 - 16 & 2476.000 & 2222.898 & 2076.541 & 2029.839 & 1955.179 & 2700.390 & 2298.766 & 2136.972 & 2036.814 & 1922.987 & 1991.302 & 2545.558 & 2545.971 & 2593.445 & 2361.087 & 2187.846 & 2221.571 & 3002.149\\

\cmidrule{1-21}
 &  & 1 - 4 & 0.089 & 0.192 & 0.177 & 0.210 & 0.171 & \textbf{\textcolor{blue}{0.062}} & 0.120 & 0.201 & 0.190 & 0.235 & 0.195 & \textbf{\textcolor{blue}{0.060}} & 0.094 & 0.192 & 0.187 & 0.224 & 0.181 & \textbf{\textcolor{blue}{0.062}}\\

 &  & 5 - 8 & 0.103 & 0.206 & 0.187 & 0.240 & 0.226 & \textbf{\textcolor{blue}{0.096}} & 0.163 & 0.216 & 0.205 & 0.272 & 0.255 & \textbf{\textcolor{blue}{0.080}} & 0.125 & 0.205 & 0.197 & 0.262 & 0.235 & \textbf{\textcolor{blue}{0.090}}\\

 & \multirow{-3}{*}{\raggedright\arraybackslash RMSCE} & 9 - 16 & 0.147 & 0.210 & 0.188 & 0.263 & 0.272 & \textbf{\textcolor{blue}{0.116}} & 0.203 & 0.236 & 0.223 & 0.305 & 0.301 & \textbf{\textcolor{blue}{0.103}} & 0.168 & 0.222 & 0.209 & 0.282 & 0.274 & \textbf{\textcolor{blue}{0.118}}\\

 &  & 1 - 4 & 0.822 & 0.804 & 0.805 & 0.781 & 0.780 & 0.803 & 0.739 & 0.798 & 0.800 & 0.760 & 0.768 & 0.810 & 0.810 & 0.811 & 0.810 & 0.772 & 0.788 & 0.811\\

 &  & 5 - 8 & 0.755 & 0.803 & 0.808 & 0.765 & 0.742 & 0.792 & 0.690 & 0.797 & 0.800 & 0.733 & 0.718 & 0.803 & 0.790 & 0.813 & 0.813 & 0.754 & 0.749 & 0.805\\

 & \multirow{-3}{*}{\raggedright\arraybackslash MC} & 9 - 16 & 0.704 & 0.803 & 0.807 & 0.747 & 0.682 & 0.788 & 0.684 & 0.797 & 0.802 & 0.716 & 0.675 & 0.798 & 0.781 & 0.812 & 0.815 & 0.751 & 0.711 & 0.804\\

 &  & 1 - 4 & 1056.000 & 1226.021 & 1510.577 & 1226.970 & 1169.282 & 937.646 & 735.943 & 1201.905 & 1440.676 & 1127.043 & 1085.666 & 828.745 & 872.429 & 1204.575 & 1411.243 & 1148.065 & 1199.830 & 760.583\\

 &  & 5 - 8 & 1552.000 & 2263.809 & 2610.437 & 1989.769 & 1715.788 & 1558.600 & 1272.274 & 2157.600 & 2744.042 & 1964.835 & 1719.941 & 1545.211 & 1489.901 & 2066.674 & 2617.772 & 2131.409 & 1882.753 & 1340.173\\

\multirow{-9}{*}{\raggedright\arraybackslash Colorado River} & \multirow{-3}{*}{\raggedright\arraybackslash Width} & 9 - 16 & 1952.000 & 3320.557 & 4295.378 & 3042.487 & 2135.137 & 2185.078 & 1723.225 & 3088.864 & 4669.426 & 2521.626 & 1863.671 & 2032.888 & 2113.742 & 2591.209 & 4371.034 & 3532.968 & 2085.335 & 1792.009\\

\cmidrule{1-21}
 &  & 1 - 4 & 0.080 & 0.154 & 0.119 & 0.159 & 0.133 & \textbf{\textcolor{blue}{0.057}} & 0.082 & 0.151 & 0.118 & 0.162 & 0.126 & \textbf{\textcolor{blue}{0.052}} & 0.103 & 0.161 & 0.126 & 0.171 & 0.136 & \textbf{\textcolor{blue}{0.048}}\\

 &  & 5 - 8 & 0.081 & 0.171 & 0.127 & 0.184 & 0.153 & \textbf{\textcolor{blue}{0.065}} & 0.105 & 0.174 & 0.128 & 0.189 & 0.148 & \textbf{\textcolor{blue}{0.069}} & 0.138 & 0.172 & 0.140 & 0.186 & 0.158 & \textbf{\textcolor{blue}{0.053}}\\

 & \multirow{-3}{*}{\raggedright\arraybackslash RMSCE} & 9 - 16 & 0.085 & 0.173 & 0.125 & 0.193 & 0.165 & \textbf{\textcolor{blue}{0.076}} & 0.116 & 0.177 & 0.135 & 0.199 & 0.163 & \textbf{\textcolor{blue}{0.085}} & 0.167 & 0.180 & 0.151 & 0.202 & 0.187 & \textbf{\textcolor{blue}{0.062}}\\

 &  & 1 - 4 & 0.762 & 0.795 & 0.799 & 0.783 & 0.777 & 0.800 & 0.779 & 0.802 & 0.803 & 0.794 & 0.783 & 0.799 & 0.721 & 0.789 & 0.791 & 0.770 & 0.765 & 0.791\\

 &  & 5 - 8 & 0.761 & 0.791 & 0.796 & 0.772 & 0.748 & 0.797 & 0.775 & 0.801 & 0.801 & 0.779 & 0.763 & 0.796 & 0.685 & 0.787 & 0.786 & 0.757 & 0.733 & 0.788\\

 & \multirow{-3}{*}{\raggedright\arraybackslash MC} & 9 - 16 & 0.771 & 0.788 & 0.793 & 0.754 & 0.714 & 0.791 & 0.792 & 0.799 & 0.798 & 0.771 & 0.732 & 0.800 & 0.664 & 0.787 & 0.788 & 0.744 & 0.699 & 0.800\\

 &  & 1 - 4 & 56320.000 & 76404.939 & 73827.493 & 75552.454 & 70955.171 & 62636.853 & 59617.219 & 76552.150 & 72739.310 & 76639.268 & 70613.688 & 62733.043 & 49136.984 & 73913.372 & 71544.580 & 72233.228 & 67414.467 & 58603.357\\

 &  & 5 - 8 & 71680.000 & 93167.307 & 89447.215 & 89840.758 & 79821.632 & 75689.833 & 76067.398 & 96834.728 & 87201.223 & 92668.130 & 79522.924 & 74593.406 & 55503.969 & 86414.134 & 83612.539 & 85192.577 & 74850.554 & 70890.704\\

\multirow{-9}{*}{\raggedright\arraybackslash Electricity Demand} & \multirow{-3}{*}{\raggedright\arraybackslash Width} & 9 - 16 & 87040.000 & 106781.678 & 105105.546 & 101789.683 & 84698.225 & 88961.105 & 91251.031 & 103152.068 & 96090.745 & 102627.903 & 84138.197 & 84352.621 & 60147.188 & 88704.532 & 87136.970 & 91908.197 & 76299.070 & 79729.275\\

\cmidrule{1-21}
 &  & 1 - 4 & 0.080 & 0.157 & 0.102 & 0.173 & 0.159 & \textbf{\textcolor{blue}{0.064}} & 0.075 & 0.152 & 0.096 & 0.170 & 0.158 & \textbf{\textcolor{blue}{0.064}} & 0.089 & 0.160 & 0.099 & 0.184 & 0.183 & \textbf{\textcolor{blue}{0.063}}\\

 &  & 5 - 8 & 0.091 & 0.168 & 0.112 & 0.192 & 0.184 & \textbf{\textcolor{blue}{0.079}} & 0.089 & 0.160 & 0.107 & 0.183 & 0.174 & \textbf{\textcolor{blue}{0.080}} & 0.086 & 0.170 & 0.107 & 0.205 & 0.203 & \textbf{\textcolor{blue}{0.078}}\\

 & \multirow{-3}{*}{\raggedright\arraybackslash RMSCE} & 9 - 16 & \textbf{\textcolor{blue}{0.097}} & 0.174 & 0.117 & 0.199 & 0.191 & 0.098 & \textbf{\textcolor{blue}{0.096}} & 0.170 & 0.113 & 0.191 & 0.194 & 0.098 & 0.092 & 0.178 & 0.111 & 0.223 & 0.210 & \textbf{\textcolor{blue}{0.090}}\\

 &  & 1 - 4 & 0.798 & 0.795 & 0.800 & 0.768 & 0.762 & 0.789 & 0.785 & 0.790 & 0.796 & 0.751 & 0.750 & 0.790 & 0.759 & 0.773 & 0.791 & 0.726 & 0.717 & 0.783\\

 &  & 5 - 8 & 0.822 & 0.792 & 0.796 & 0.755 & 0.742 & 0.790 & 0.792 & 0.791 & 0.795 & 0.745 & 0.733 & 0.791 & 0.768 & 0.771 & 0.789 & 0.707 & 0.695 & 0.785\\

 & \multirow{-3}{*}{\raggedright\arraybackslash MC} & 9 - 16 & 0.833 & 0.788 & 0.794 & 0.745 & 0.726 & 0.782 & 0.794 & 0.789 & 0.796 & 0.737 & 0.714 & 0.786 & 0.767 & 0.766 & 0.789 & 0.686 & 0.682 & 0.781\\

 &  & 1 - 4 & 12.734 & 14.549 & 13.906 & 14.001 & 13.394 & 12.739 & 13.010 & 14.892 & 14.588 & 13.912 & 13.357 & 13.626 & 12.815 & 14.556 & 15.296 & 13.468 & 13.126 & 13.895\\

 &  & 5 - 8 & 14.473 & 15.482 & 14.649 & 14.663 & 13.861 & 13.378 & 14.156 & 15.832 & 15.307 & 14.792 & 14.048 & 14.229 & 14.801 & 15.603 & 16.389 & 14.394 & 13.818 & 15.098\\

\multirow{-9}{*}{\raggedright\arraybackslash NO2 Concentration} & \multirow{-3}{*}{\raggedright\arraybackslash Width} & 9 - 16 & 15.367 & 15.585 & 15.415 & 14.712 & 13.907 & 13.694 & 14.682 & 16.332 & 15.968 & 14.852 & 14.236 & 14.423 & 15.253 & 16.000 & 17.375 & 14.544 & 14.028 & 15.538\\
\bottomrule
\end{tabular}}
\endgroup{}
\end{table}
\end{landscape}
We now examine whether this first-step adaptivity advantage translates into improved local empirical coverage in finite-sample forecasting experiments. To do so, we compare interval performance across four real-world time-series datasets: bike rental counts in the Washington metropolitan area from Capital Bikeshare~\citep{capital_bikeshare_data}, electricity demand in New York State from the New York Independent System Operator (NYISO)~\citep{nyis_electricity}, NO$_2$ concentration in Pasadena, CA (EPA AQS site 06-037-2005)~\citep{epa}, and streamflow (discharge in cubic feet per second) of the Colorado River near Cisco, UT (USGS NWIS site 09180500, parameter 00060)~\citep{colorado}. The corresponding time-series plots are provided in Figure~\ref{fig:data} of Appendix~\ref{app:addf}.

All methods are evaluated under a rolling-origin forecasting protocol. Each TSFM receives a context window of $256$ consecutive observations and produces predictive quantiles at horizons $h=1,\ldots,H$, with $H=16$, at the decile levels $\{10\%,20\%,\ldots,90\%\}$. The forecast origin advances one step at a time through the test period. Because the widest central prediction interval directly available from these deciles spans the 10th to the 90th percentile, corresponding to 80\% nominal coverage, we use $\alpha=0.2$ as the primary evaluation level. We compare prediction intervals from the native forecaster, VCPI, ACI~\citep{conformal2}, QT~\citep{conformal_pid}, and SPCI~\citep{spci}. Comparisons with EnbPI~\citep{enbpi} are also made but are not included in the main text, as SPCI, an extension of EnbPI, performs strictly better. For the conformal baselines, ACI uses a learning rate of $\gamma=0.005$, while QT uses a learning rate equal to $0.05$ times the mean absolute residual in the calibration window, following the conventions used in their respective papers~\citep{conformal2,conformal_pid}. Neither paper provides explicit guidance for learning-rate selection in new settings. We discuss sensitivity to these choices in Tables~\ref{tab:aci_sensitivity} and~\ref{tab:qt_sensitivity} of Appendix~\ref{app:addf}. Full implementation details for each method are provided in Appendix~\ref{app:implementation}.

We compare uncertainty quantification methods using three complementary metrics: marginal coverage (MC), root mean squared coverage error (RMSCE), and median interval width. MC measures the proportion of observations falling within their prediction intervals and therefore assesses overall calibration, while median width reflects interval efficiency through the typical size of the prediction intervals. Our primary focus is RMSCE, which is defined by first computing rolling empirical coverage over a window of length $\ell$ and then measuring its root mean squared deviation from the target level $1-\alpha$. For a fixed horizon $h$ and miscoverage level $\alpha$, let $\widehat C_{u,\alpha}(\mathcal F_{u-h})$ denote a method's prediction interval for $Y_u$, constructed using information available at time $u-h$. Define
\begin{align*} 
	\widehat{\mathrm{MC}}_{t,h}^{(\ell)} &= \frac{1}{\ell}\sum_{u=t-\ell+1}^{t} \mathbf 1\{Y_u\in\widehat C_{u,\alpha}(\mathcal F_{u-h})\}, & \mathrm{RMSCE}_h &= \sqrt{\frac{1}{|\mathcal T_{\ell,h}|}  \sum_{t\in\mathcal T_{\ell,h}} \left(\widehat{\mathrm{MC}}_{t,h}^{(\ell)}-(1-\alpha)\right)^2},
\end{align*}
where $\ell$ is the evaluation-window length, $\mathcal T_{\ell,h}$ is the set of endpoints of valid rolling windows containing $\ell$ consecutive evaluated targets at horizon $h$, and $\mathbf 1\{\cdot\}$ denotes the indicator function. In all experiments, we set $\ell=100$. 

Unlike MC, which averages over the full evaluation period, RMSCE measures how well a method maintains calibration locally over time. This distinction is central in our setting because a method can achieve the correct average coverage while still exhibiting substantial undercoverage or overcoverage during periods of changing uncertainty. RMSCE is a diagnostic of local empirical coverage, not an estimator or guarantee of coverage conditional on the full information set. We therefore use it to assess local coverage stability; definitions of the remaining metrics are provided in Appendix~\ref{app:eval_details}.

Table~\ref{tab:cov_80} summarizes the results of our experiments. We evaluate forecasts at horizons $h=1,\ldots,16$ and aggregate the results into three horizon blocks: 1--4, 5--8, and 9--16. The clearest pattern in Table~\ref{tab:cov_80} concerns RMSCE. VCPI attains the smallest RMSCE in nearly every dataset, model, and horizon-block combination, indicating substantially improved local calibration under time-varying uncertainty. Among conformal methods, SPCI achieves the best RMSCE by fitting a quantile random forest to lagged residuals, yet still falls substantially short of VCPI across most settings. ACI and QT exhibit the largest RMSCE values, and the native intervals, despite achieving broadly reasonable marginal coverage, show substantial local miscalibration. A visual comparison of local calibration is also provided in Figure~\ref{fig:result2} of Appendix~\ref{app:addf}.

In terms of MC, VCPI tends to overcover slightly relative to the nominal 80\% level. We do not view this as a deficiency. Through a bias--variance lens, low RMSCE indicates that rolling coverage tracks the nominal level closely over time, meaning that the intervals are \emph{locally} well calibrated, with low variance around the target. The mild positive bias in marginal coverage is a small price for this stability. Crucially, this overcoverage does not come at the cost of uninformatively wide intervals: on Colorado River and Electricity Demand, VCPI produces intervals that are competitive with or narrower than the native intervals and substantially narrower than those from ACI and QT. SPCI produces the narrowest intervals on NO$_2$ Concentration and Electricity Demand, but achieves this by undercovering relative to the nominal level, resulting in larger RMSCE than VCPI. On Bike Rental Counts, SPCI also produces narrower intervals than VCPI at longer horizons, again at the cost of worse RMSCE, illustrating the expected trade-off between interval tightness and local calibration.

This pattern is not specific to the $\alpha=0.2$ setting. The corresponding results for $\alpha=0.1$ and $\alpha=0.05$, reported in Tables~\ref{tab:cov_90} and~\ref{tab:cov_95} of the appendix, are broadly consistent: VCPI maintains a clear RMSCE advantage across datasets and models at the 90\% level and continues to dominate at short horizons (1--4 steps) at the 95\% level. Its advantage is less pronounced at longer horizons, particularly on Colorado River at 95\% nominal coverage, where ACI becomes competitive. We attribute this to the well-known mean reversion of multi-step GARCH forecasts, combined with increased sensitivity to distributional misspecification near the tails, a limitation we discuss further in Section~\ref{sec:conc}.

We also conduct sensitivity analyses for ACI and QT, both of which require a learning-rate choice. We repeat the analysis using several alternative learning rates for each method. The results, summarized in Tables~\ref{tab:aci_sensitivity} and~\ref{tab:qt_sensitivity}, show some sensitivity to this choice. Even under the most favorable learning-rate choice for each method, neither ACI nor QT consistently approaches the RMSCE achieved by VCPI.

\section{Conclusion}\label{sec:conc}
This paper identifies local scale miscalibration as a practically important failure mode of probabilistic time-series forecasting. Even when prediction intervals achieve near-nominal marginal coverage, their local reliability can vary substantially over time as forecast uncertainty changes across regimes. Our empirical results show that the native uncertainty estimates produced by current TSFMs are often not sufficient on their own: although their marginal coverage can appear reasonable, their local calibration is frequently unstable, indicating the need for additional post-hoc correction. We propose VCPI, a post-hoc calibration method that preserves the forecaster's predictive location and baseline scale while correcting residual volatility through a multiplicative GARCH-based adjustment. Our theoretical comparison concerns the first prediction after observing the initial post-break residual, conditional on a specified shock event and under the stated assumptions on GARCH and QT. Empirically, across three TSFMs and four real-world datasets, VCPI consistently yields the smallest RMSCE, showing an advantage in local empirical coverage stability, while remaining broadly competitive in marginal coverage and interval width.

\paragraph{Limitations} VCPI relies on a parametric GARCH-family specification for the residual scale and a parametric standardized error distribution $F$ estimated from normalized residuals. When either the scale dynamics or the standardized error distribution is misspecified, the resulting intervals may inherit specification bias. Both choices trade flexibility for the first-step adaptivity property: when the residual scale process is poorly approximated by GARCH or when the predictive quantiles are inconsistent with any element of the reference family, the resulting intervals may inherit specification bias. This is most visible at longer horizons under high nominal coverage: multi-step GARCH forecasts mean-revert to the unconditional variance relatively quickly, degrading the volatility correction at horizons beyond 4--8 steps, and tail coverage at the 95\% level is more sensitive to distributional misspecification than moderate coverage levels, compounding the effect. Nevertheless, VCPI remains competitive across the evaluated settings, even where it does not achieve the lowest RMSCE.
\newpage
\bibliographystyle{plainnat}
\bibliography{ref}
\newpage

\appendix
\section{VCPI Algorithm for Multi-Step Prediction}\label{app:vcpi_algorithm}
Algorithm~\ref{alg:vcpi_overall} describes VCPI at forecast origin $t$, given the information $\mathcal F_t$ available through time $t$. Let $H$ be the maximum forecast horizon, $W$ the calibration-window length, and $\mathcal A\subset(0,1)$ the set of nominal miscoverage levels. For each horizon $h=1,\ldots,H$, we fit a separate model to
\[
e_u^{(h)}
=
\frac{Y_u-\hat f_u(\mathcal F_{u-h})}
{\hat\sigma_u(\mathcal F_{u-h})},
\qquad u=t-W+1,\ldots,t,
\]
where $\hat f_u(\mathcal F_{u-h})$ is the $h$-step-ahead point forecast of $Y_u$, and $\hat\sigma_u(\mathcal F_{u-h})$ is the scale extracted from its predictive quantiles at levels $p_1,\ldots,p_K$ using Equation~\eqref{eq:quant_optim}.

Let $\mathcal S$ denote the candidate pairs of variance specification and standardized error distribution. In our experiments, these are standard GARCH, GJR-GARCH, and EGARCH, each of order $(1,1)$, paired with normal or standardized skewed-$t$ errors. We use $W=256$ and split each residual sequence chronologically into 80\% training and 20\% validation, indexed by $\mathcal D_{\mathrm{train}}$ and $\mathcal D_{\mathrm{val}}$, respectively. For candidate $S$, let $\widehat C^{S}_{u,h,\alpha}$ denote its validation prediction interval for $Y_u$ at nominal coverage $1-\alpha$. We select the candidate minimizing
\[ L_v(S) = \frac{1}{|\mathcal A|} \sum_{\alpha\in\mathcal A} \left| \frac{1}{|\mathcal D_{\mathrm{val}}|} \sum_{u\in\mathcal D_{\mathrm{val}}} \mathbf 1\{Y_u\in\widehat C^{S}_{u,h,\alpha}\} -(1-\alpha) \right|,
\]
the absolute validation coverage error averaged over $\mathcal A$, where $\mathbf 1\{\cdot\}$ denotes the indicator function. After refitting on the full window, the selected horizon-$h$ model provides the $h$-step scale forecast $\tilde s_{t+h}^{(h)}(\mathcal F_t)$ and fitted standardized error distribution $\tilde F_t^{(h)}$ used to construct the interval for $Y_{t+h}$.

For each horizon $h$, the corrected interval is
\begin{equation}\label{eq:vcpi_multihorizon_interval}
\begin{aligned}
\widehat C_{t+h,\alpha}^{\mathrm{VCPI}}(\mathcal F_t)
=\big[&
\hat f_{t+h}(\mathcal F_t)
+\hat\sigma_{t+h}(\mathcal F_t)
\tilde s_{t+h}^{(h)}(\mathcal F_t)
(\tilde F_t^{(h)})^{-1}(\alpha/2),\\
&
\hat f_{t+h}(\mathcal F_t)
+\hat\sigma_{t+h}(\mathcal F_t)
\tilde s_{t+h}^{(h)}(\mathcal F_t)
(\tilde F_t^{(h)})^{-1}(1-\alpha/2)
\big].
\end{aligned}
\end{equation}
\begin{algorithm}[h]
\caption{VCPI at forecast origin $t$}\label{alg:vcpi_overall}
\begin{algorithmic}[1]
\Require Observations through $t$; forecasts and predictive quantiles; maximum horizon $H$; window length $W$; quantile levels $p_1,\ldots,p_K$; miscoverage levels $\mathcal A$; candidate specifications $\mathcal S$.
\Ensure Prediction intervals $\widehat C_{t+h,\alpha}^{\mathrm{VCPI}}(\mathcal F_t)$ for $Y_{t+h}$ at nominal coverage $1-\alpha$, for $h=1,\ldots,H$ and $\alpha\in\mathcal A$.
\For{$h=1,\ldots,H$}
  \State Collect the observations and corresponding $h$-step-ahead forecasts
  $\{(Y_u,\hat f_u(\mathcal F_{u-h}))\}_{u=t-W+1}^{t}$.
  \State Compute $\hat\sigma_u(\mathcal F_{u-h})$ using Equation~\eqref{eq:quant_optim}.
  \State Form $e_u^{(h)}=\{Y_u-\hat f_u(\mathcal F_{u-h})\}/\hat\sigma_u(\mathcal F_{u-h})$.
  \State Split $\{e_{u}^{(h)}\}_{u=t-W+1}^{t}$ chronologically into training and validation indices $\mathcal D_{\mathrm{train}}$ and $\mathcal D_{\mathrm{val}}$.
  \For{$S\in\mathcal S$}
    \State Fit $S$ on $\{e_u^{(h)}:u\in\mathcal D_{\mathrm{train}}\}$ and evaluate $L_v(S)$ on $\mathcal D_{\mathrm{val}}$.
  \EndFor
  \State Select $S^\star=\arg\min_{S\in\mathcal S}L_v(S)$.
  \State Refit $S^\star$ on $\{e_u^{(h)}\}_{u=t-W+1}^{t}$ and forecast the residual scale $h$ steps ahead.
  \State Compute $\hat\sigma_{t+h}(\mathcal F_t)$ from the current horizon-$h$ predictive quantiles using Equation~\eqref{eq:quant_optim}.
  \For{$\alpha\in\mathcal A$}
    \State Construct $\widehat C_{t+h,\alpha}^{\mathrm{VCPI}}(\mathcal F_t)$ using Equation~\eqref{eq:vcpi_multihorizon_interval}.
  \EndFor
\EndFor
\end{algorithmic}
\end{algorithm}

\newpage
\section{Proofs}
\subsection{Asymptotic upper bound for GARCH width}\label{app:lem1}
Let $(\hat\omega,\hat a,\hat b)$ denote the GARCH(1,1) coefficients estimated using $\mathcal F_\tau$, including $e_\tau$. Let $\hat h_\tau$ denote the fitted variance state at time $\tau$ computed using these coefficients. Fix $\rho\in(0,1)$ and consider the event $E_\tau^{\mathrm{shock}}(\rho)$.
Using the scale-forecast notation from Section~\ref{sec:method}, 
\[
\hat h_{\tau+1}(\mathcal F_\tau) =\tilde s_{\tau+1}^{\,2}(\mathcal F_\tau).
\]
The corresponding one-step-ahead half-width is
\[
\hat q^G_{\tau+1}(\mathcal F_\tau) =c_\alpha\sqrt{\hat h_{\tau+1}(\mathcal F_\tau)}.
\]
Define
\[
\bar h_\tau = \frac{\hat\omega}{1-\hat a-\hat b},
\]
the unconditional variance implied by the fitted GARCH model.

The GARCH recursion gives the following decomposition:
\[
\begin{aligned}
\hat h_{\tau+1}(\mathcal F_\tau)-s_{\tau+1}^2
={}&
\underbrace{\hat a(e_\tau^2-s_{\tau+1}^2)}
_{\text{Shock mismatch}}
+
\underbrace{(1-\hat a)(\bar h_\tau-s_{\tau+1}^2)}
_{\text{Long-run level mismatch}} +\underbrace{\hat b(\hat h_\tau-\bar h_\tau)}
_{\text{State deviation}}.
\end{aligned}
\]
The three terms compare the observed squared residual with the oracle variance, the fitted model's long-run variance with the oracle variance, and the fitted variance state with the model's long-run variance, respectively. 

By the triangle inequality,
\[
\begin{aligned}
\bigl|\hat h_{\tau+1}(\mathcal F_\tau)-s_{\tau+1}^2\bigr|
\le &
\hat a\bigl|e_\tau^2-s_{\tau+1}^2\bigr| +(1-\hat a)\bigl|\bar h_\tau-s_{\tau+1}^2\bigr| +\hat b\bigl|\hat h_\tau-\bar h_\tau\bigr|.
\end{aligned}
\]
Since $|x-y|=|x^2-y^2|/(x+y)$ for $x\ge0$ and $y>0$,
\[
\begin{aligned}
\bigl|\hat q^G_{\tau+1}(\mathcal F_\tau)-q^\star_{\tau+1}\bigr| &=
\frac{c_\alpha \bigl|\hat h_{\tau+1}(\mathcal F_\tau)-s_{\tau+1}^2\bigr|}{\sqrt{\hat h_{\tau+1}(\mathcal F_\tau)}+s_{\tau+1}} 
\le \frac{c_\alpha}{s_{\tau+1}} \bigl|\hat h_{\tau+1}(\mathcal F_\tau)-s_{\tau+1}^2\bigr|.
\end{aligned}
\]
Substituting $s_{\tau+1}=v(\tau+1)+\Delta$ gives
\begin{equation*}
\begin{aligned}
\bigl|\hat q^G_{\tau+1}(\mathcal F_\tau)-q^\star_{\tau+1}\bigr|
\le{}&
\frac{c_\alpha}{v(\tau+1)+\Delta}
\Bigl[
\hat a\bigl|e_\tau^2-(v(\tau+1)+\Delta)^2\bigr|\\
&\quad +(1-\hat a)
\bigl|\bar h_\tau-(v(\tau+1)+\Delta)^2\bigr|
+\hat b\bigl|\hat h_\tau-\bar h_\tau\bigr|
\Bigr].
\end{aligned}
\end{equation*}

\paragraph{Shock Mismatch}
On the event $E_\tau^{\mathrm{shock}}(\rho)$,
\[
\bigl|e_\tau^2-(v(\tau)+\Delta)^2\bigr| \le \rho(v(\tau)+\Delta)^2.
\]
By Assumption~\ref{as:vt},
$|v(\tau+1)-v(\tau)|\le\eta_{\tau,W}$. Therefore,
\begin{align*}
\bigl|e_\tau^2-(v(\tau+1)+\Delta)^2\bigr| &\le \bigl|e_\tau^2-(v(\tau)+\Delta)^2\bigr| + \bigl|(v(\tau)+\Delta)^2-(v(\tau+1)+\Delta)^2\bigr|\\
&\le \rho(v(\tau)+\Delta)^2 + \bigl|v(\tau)^2-v(\tau+1)^2 +2\Delta(v(\tau)-v(\tau+1))\bigr|\\
&\le \rho(v(\tau)+\Delta)^2 +\eta_{\tau,W}(v(\tau)+v(\tau+1)) +2\Delta\eta_{\tau,W}\\
&= \rho(v(\tau)+\Delta)^2 +\eta_{\tau,W}\bigl(v(\tau)+v(\tau+1)+2\Delta\bigr).
\end{align*}

\paragraph{State Deviation.}
We show that $|\hat h_\tau-\bar h_\tau|=O_p(1)$. By Assumption~\ref{as:garch}, the fitted variance recursion starts at
\[
\hat h_{\tau-W+1}=\bar h_\tau
=\frac{\hat\omega}{1-\hat a-\hat b},
\qquad
0<\bar h_\tau\le \underline\omega / \kappa.
\]
Using the same fitted coefficients throughout the window,
\[
\hat h_u=\hat\omega+\hat a e_{u-1}^2+\hat b\hat h_{u-1},
\qquad u=\tau-W+2,\ldots,\tau.
\]
Expanding this recursion gives
\[
\hat h_\tau
=
\hat b^{W-1}\bar h_\tau
+
\sum_{j=0}^{W-2}
\hat b^j\bigl(\hat\omega+\hat a e_{\tau-1-j}^2\bigr).
\]
Since $0<\hat a,\hat b<1$,
\[
0\le\hat h_\tau \le \underline\omega / \kappa +(W-1)\underline\omega +\sum_{u=\tau-W+1}^{\tau-1}e_u^2.
\]
The sum involves only pre-jump residuals, whose joint distribution
does not depend on $\Delta$. Because $W$ is fixed, this sum is
$O_p(1)$ as $\Delta\to\infty$. Hence $\hat h_\tau=O_p(1)$, and
\[
|\hat h_\tau-\bar h_\tau|
\le\hat h_\tau+\bar h_\tau
=O_p(1).
\]
This bound holds even though the fitted coefficients depend on
$e_\tau$, because their bounds are uniform in $\Delta$.

For fixed $\rho\in(0,1)$, the event
$E_\tau^{\mathrm{shock}}(\rho)=\{|\xi_\tau^2-1|\le\rho\}$
has positive probability independent of $\Delta$. Therefore,
the same $O_p(1)$ bound holds conditional on this event. Combining these derivations, we get
\begin{align*}
\bigl|\hat q^G_{\tau+1}(\mathcal F_\tau)-q_{\tau+1}^\star\bigr|
\le\;&
\frac{c_\alpha}{v(\tau+1)+\Delta}
\Bigl[
\hat a\Bigl\{
\rho (v(\tau)+\Delta)^2
+
\eta_{\tau,W}\bigl(v(\tau)+v(\tau+1)+2\Delta\bigr)
\Bigr\}\\
&\qquad
+
(1-\hat a)\Bigl|
\bar{h}_{\tau} - (v(\tau+1)+\Delta)^2
\Bigr| + \hat b\,|\hat h_{\tau} - \bar h_{\tau}|
\Bigr] \\
\le\;&
\frac{c_\alpha}{v(\tau+1)+\Delta}
\Bigl[
\hat a\Bigl\{
\rho (v(\tau)+\Delta)^2
+
\eta_{\tau,W}\bigl(v(\tau)+v(\tau+1)+2\Delta\bigr)
\Bigr\}\\
&\qquad
+
(1-\hat a)\Bigl\{
\bar{h}_{\tau}+(v(\tau+1)+\Delta)^2
\Bigr\} + \hat b\,|\hat h_{\tau} - \bar h_{\tau}|
\Bigr] \\
\le\;&
\frac{c_\alpha}{v(\tau+1)+\Delta}
\Bigl[\hat a \rho v(\tau)^2+ \hat a \rho \Delta^2
+ 2 \hat a \rho v(\tau)\Delta\\
&\qquad + \hat a \eta_{\tau,W}v(\tau)
+\hat a \eta_{\tau,W}v(\tau+1)
+2 \hat a \Delta \eta_{\tau,W}\\
&\qquad + (1-\hat a)\bar{h}_{\tau}
+(1-\hat a)v(\tau+1)^2+(1-\hat a)\Delta^2\\
&\qquad + 2(1-\hat a)v(\tau+1)\Delta
+ \hat b\,|\hat h_{\tau} - \bar h_{\tau}| \Bigr]\\
\le\;&
\frac{c_\alpha}{v(\tau+1)+\Delta}
\Bigl[\hat a \rho v(\tau)^2
+ \hat a \eta_{\tau,W}v(\tau)
+ \hat a \eta_{\tau,W}v(\tau+1)
+ (1-\hat a)\bar{h}_{\tau}
+ (1-\hat a)v(\tau+1)^2\\
&\qquad + (1-\hat a(1-\rho))\Delta^2\\
&\qquad + 2\hat a\rho v(\tau)\Delta
+ 2\hat a\Delta\eta_{\tau,W}\\
&\qquad + 2(1-\hat a)v(\tau+1)\Delta
+ \hat b\,|\hat h_{\tau}-\bar h_{\tau}| \Bigr]\\
\le\;&
\frac{c_\alpha}{v(\tau+1)+\Delta}
\Bigl[(1-\underline a)
\Bigl(\rho v(\tau)^2
+ \eta_{\tau,W}v(\tau)
+ \eta_{\tau,W}v(\tau+1)
+ \frac{\underline\omega}{\kappa}
+ v(\tau+1)^2\Bigr)\\
&\qquad + (1-\underline a(1-\rho))\Delta^2\\
&\qquad + (1-\underline a)
\bigl(2\rho v(\tau)\Delta
+ 2\Delta\eta_{\tau,W}
+ 2v(\tau+1)\Delta\bigr)
+ (1-\underline b)\,|\hat h_{\tau}-\bar h_{\tau}| \Bigr]\\
\le\;&
\frac{c_\alpha}{v(\tau+1)+\Delta}
\Bigl[
\bigl(1-\underline a(1-\rho)\bigr)\Delta^2
+O(\Delta)+O_p(1)
\Bigr].
\end{align*}
The last line follows from
$\bigl|\hat h_\tau-\bar h_\tau\bigr|=O_p(1)$.

Since
\[
\frac{\Delta^2}{v(\tau+1)+\Delta}=\Delta+O(1),
\qquad
\frac{O(\Delta)}{v(\tau+1)+\Delta}=O(1),
\qquad
\frac{O_p(1)}{v(\tau+1)+\Delta}=O_p(1),
\]
it follows that
\[
\bigl|\hat q^G_{\tau+1}(\mathcal F_\tau)-q_{\tau+1}^\star\bigr|
\le
c_\alpha\bigl[1-\underline{a}(1 - \rho)\bigr]\Delta+O_p(1).
\]
Because \(0<\rho<1\) and \(\underline a>0\), the coefficient \(1-\underline a(1-\rho)\) is strictly smaller than \(1\).

\subsection{Quantile Tracking}\label{app:lem2}
%

QT updates its estimated half-width according to
\[
\hat q_{t+1}^{QT}(\mathcal F_t)
=
\hat q_t^{QT}(\mathcal F_{t-1})
+\gamma(\mathrm{miss}_t-\alpha),
\]
where
\[
\mathrm{miss}_t
:=\mathbf 1\{|e_t|>\hat q_t^{QT}(\mathcal F_{t-1})\},
\]
and $\gamma>0$ is fixed independently of $\Delta$.

Suppose QT is initialized at a fixed time $t_0<\tau$ with a
finite half-width computed from pre-jump observations.
Its initial half-width is therefore $O_p(1)$ as
$\Delta\to\infty$. Since $|\mathrm{miss}_t-\alpha|\le1$,
\[
\bigl|\hat q_\tau^{QT}(\mathcal F_{\tau-1})\bigr|
\le
\bigl|\hat q_{t_0}^{QT}(\mathcal F_{t_0-1})\bigr|
+\gamma(\tau-t_0)
=O_p(1).
\]
The update after observing the shock also has magnitude at most
$\gamma$, so
\[
\bigl|\hat q_{\tau+1}^{QT}(\mathcal F_\tau)\bigr|
\le
\bigl|\hat q_\tau^{QT}(\mathcal F_{\tau-1})\bigr|+\gamma
=O_p(1).
\]
The oracle half-width is
\[
q_{\tau+1}^\star
=c_\alpha(v(\tau+1)+\Delta).
\]
By the reverse triangle inequality and $v(\tau+1)>0$,
\[
\begin{aligned}
\bigl|\hat q_{\tau+1}^{QT}(\mathcal F_\tau)-q_{\tau+1}^\star\bigr|
&\ge q_{\tau+1}^\star
-\bigl|\hat q_{\tau+1}^{QT}(\mathcal F_\tau)\bigr|\\
&\ge c_\alpha\Delta
-\bigl|\hat q_{\tau+1}^{QT}(\mathcal F_\tau)\bigr|\\
&=c_\alpha\Delta-O_p(1),
\end{aligned}
\]
as $\Delta\to\infty$.

\subsection{Asymptotic Dominance}\label{app:theorem}
By Lemma~\ref{lem:garch_asymptotic}, conditional on
$E_\tau^{\mathrm{shock}}(\rho)$,
\[
\bigl|\hat q_{\tau+1}^{G}(\mathcal F_\tau)-q^\star_{\tau+1}\bigr|
\le
c_\alpha\bigl[1-\underline a(1-\rho)\bigr]\Delta+O_p(1).
\]
By Lemma~\ref{lem:aci_lower},
\[
\bigl|\hat q_{\tau+1}^{QT}(\mathcal F_\tau)-q^\star_{\tau+1}\bigr|
\ge
c_\alpha\Delta-O_p(1).
\]
Since $E_\tau^{\mathrm{shock}}(\rho)$ has positive probability
independent of $\Delta$, the latter remainder is also $O_p(1)$
conditional on this event. Subtracting the two bounds gives
\[
\begin{aligned}
&\bigl|\hat q_{\tau+1}^{QT}(\mathcal F_\tau)-q^\star_{\tau+1}\bigr|
-
\bigl|\hat q_{\tau+1}^{G}(\mathcal F_\tau)-q^\star_{\tau+1}\bigr|\\
&\qquad\ge
c_\alpha\underline a(1-\rho)\Delta-O_p(1),
\end{aligned}
\]
under the same conditional law.

Because $c_\alpha\underline a(1-\rho)>0$, the right-hand side
diverges to $+\infty$ in conditional probability as
$\Delta\to\infty$. Hence, for any fixed $\delta>0$,
\[
\mathbb P\!\left(
\bigl|\hat q_{\tau+1}^{QT}(\mathcal F_\tau)-q^\star_{\tau+1}\bigr|
-
\bigl|\hat q_{\tau+1}^{G}(\mathcal F_\tau)-q^\star_{\tau+1}\bigr|
\ge\delta
\,\middle|\,
E_\tau^{\mathrm{shock}}(\rho)
\right)\to1.
\]

\newpage
\section{Additional Figures and Tables}\label{app:addf}
\begin{figure}[ht]
\centering
\begin{subfigure}{0.48\linewidth}
    \centering
    \includegraphics[width=\linewidth]{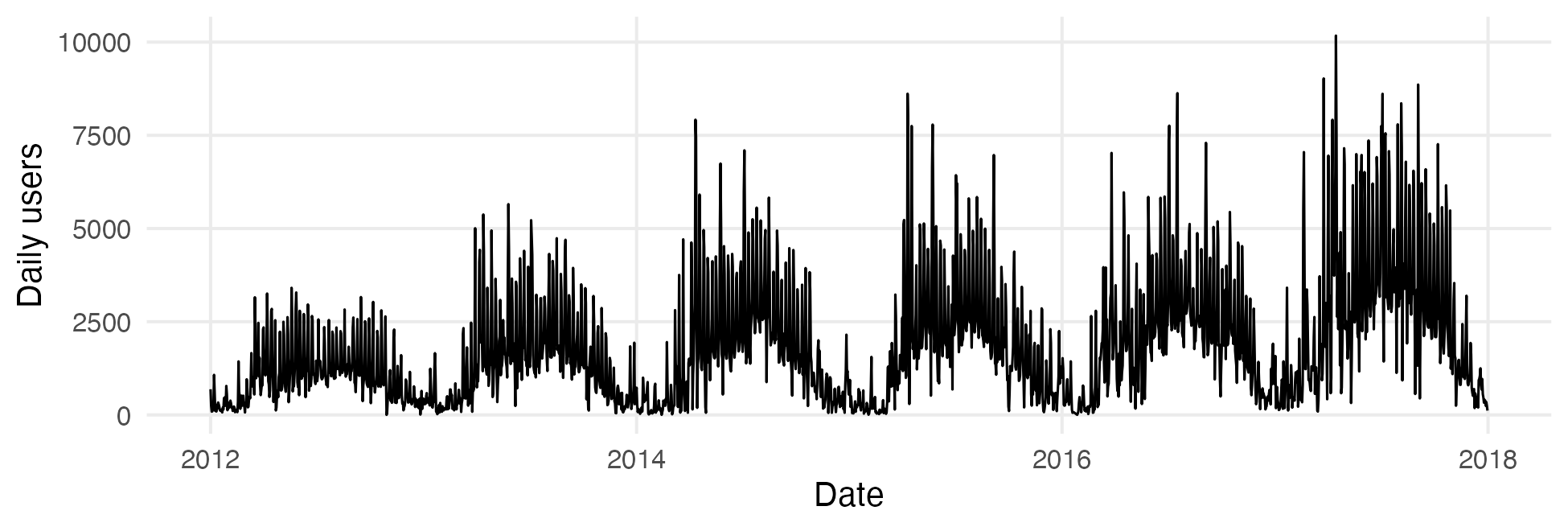}
\end{subfigure}
\hfill
\begin{subfigure}{0.48\linewidth}
    \centering
    \includegraphics[width=\linewidth]{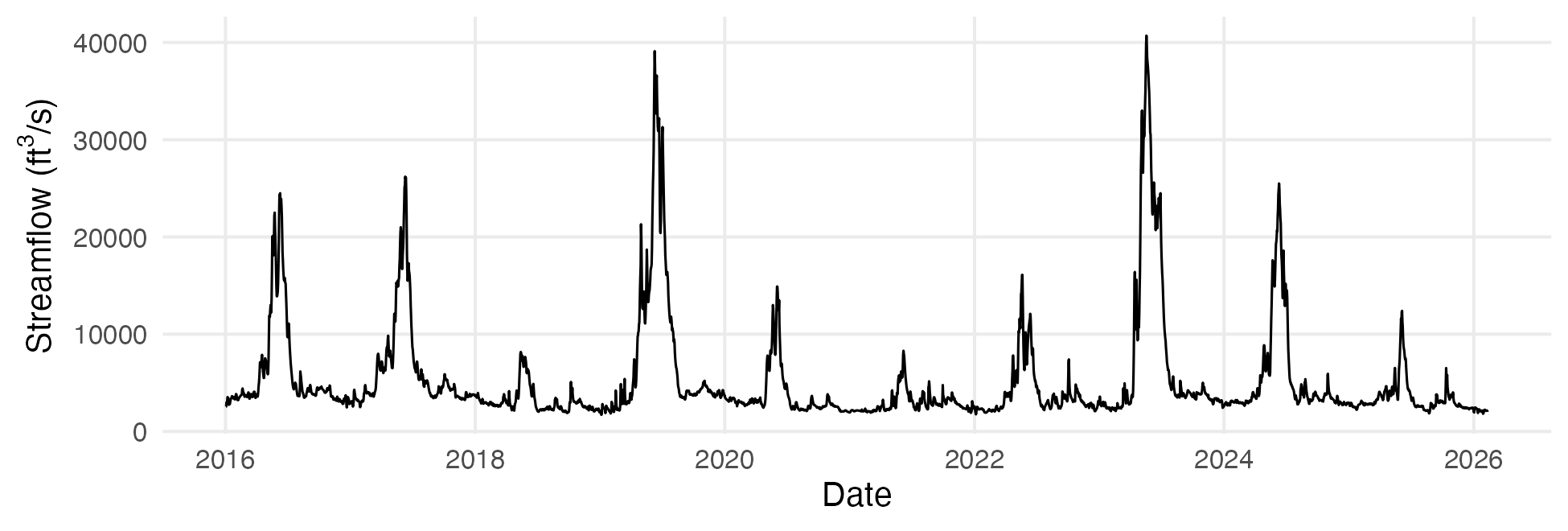}
\end{subfigure}
\begin{subfigure}{0.48\linewidth}
    \centering
    \includegraphics[width=\linewidth]{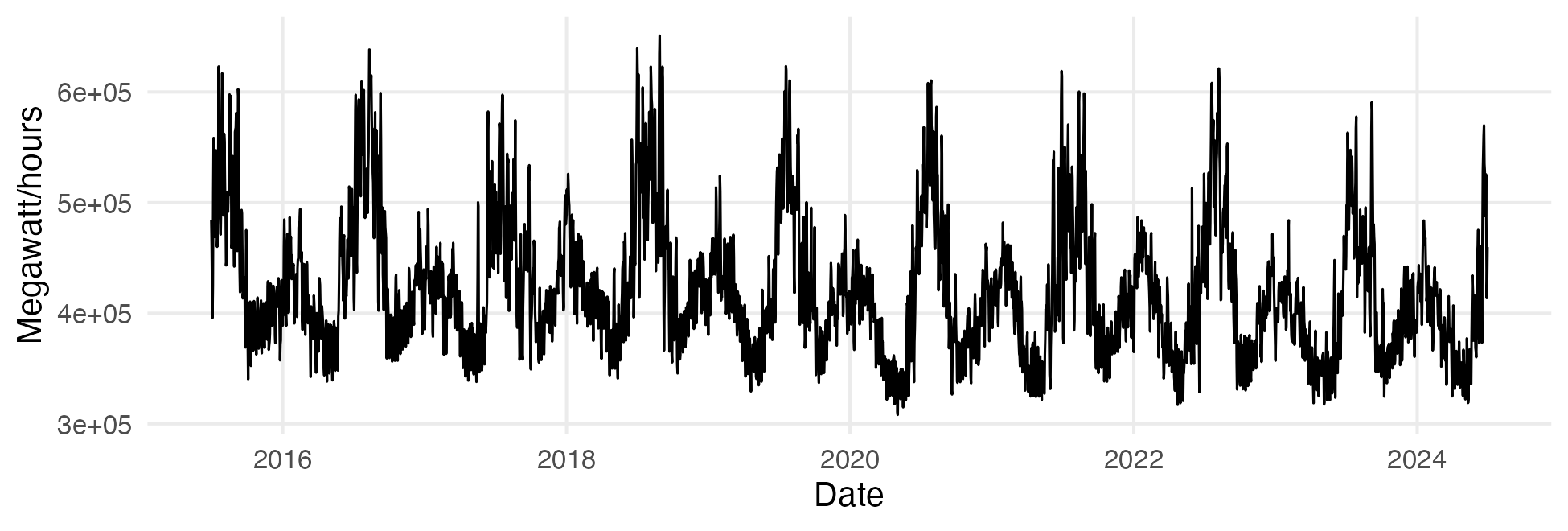}
\end{subfigure}
\hfill
\begin{subfigure}{0.48\linewidth}
    \centering
    \includegraphics[width=\linewidth]{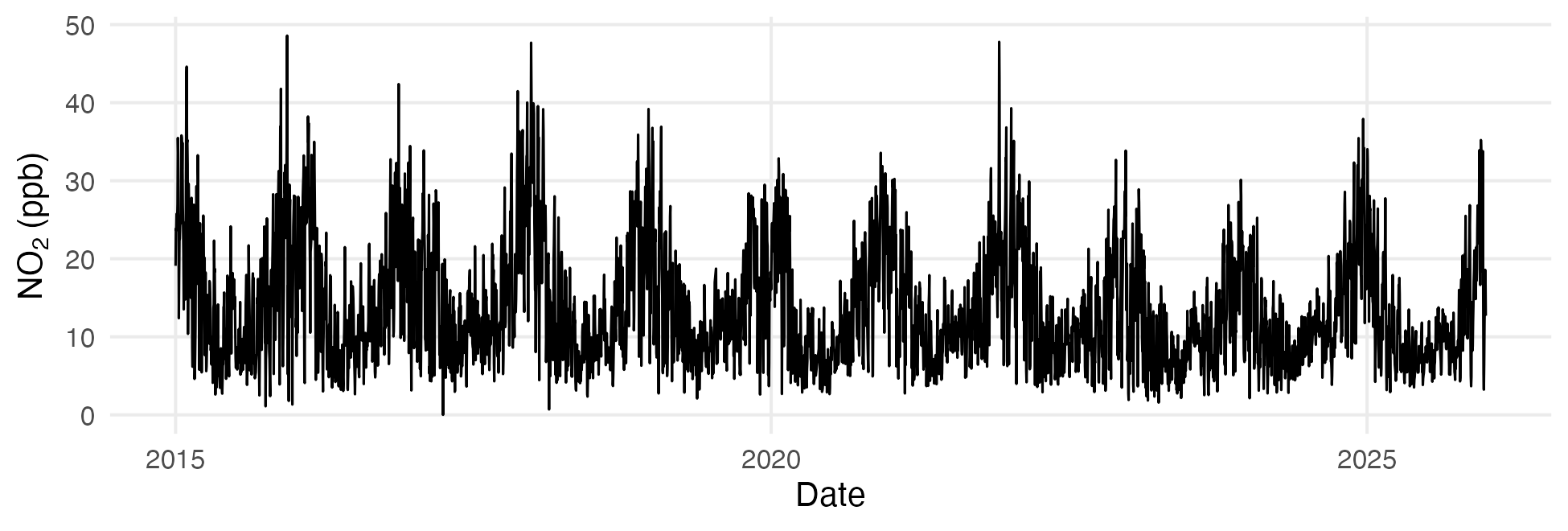}
\end{subfigure}
\caption{Four daily time-series datasets are considered: bike rental counts in the Washington metropolitan area via Capital Bikeshare~\citep{capital_bikeshare_data}, electricity demand in New York State from the New York Independent System Operator (NYISO)~\citep{nyis_electricity}, NO$_2$ concentration in Pasadena, CA (EPA AQS site 06-037-2005)~\citep{epa}, and streamflow (discharge, cubic feet per second) of the Colorado River near Cisco, UT (USGS NWIS site 09180500, parameter 00060)~\citep{colorado}. A common feature across these datasets is pronounced time-varying volatility, with distinct regimes in which the variability of the underlying process changes substantially over time. Such heteroskedastic behavior makes certain periods intrinsically more difficult to predict than others.}
\label{fig:data}
\end{figure}

\begin{figure}[ht]
\centering
\begin{subfigure}{0.48\linewidth}
    \centering
    \includegraphics[width=\linewidth]{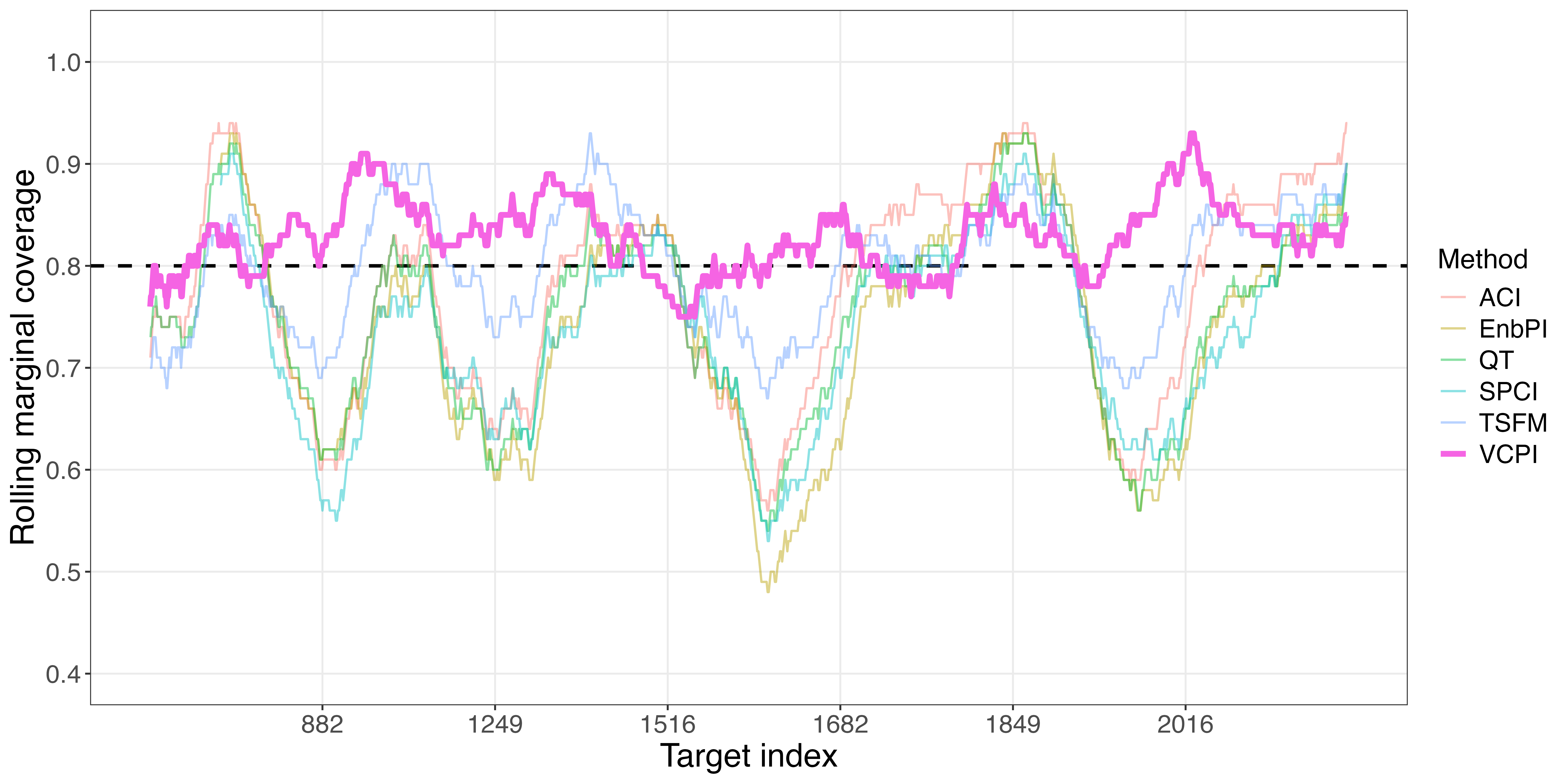}
\end{subfigure}
\hfill
\begin{subfigure}{0.48\linewidth}
    \centering
    \includegraphics[width=\linewidth]{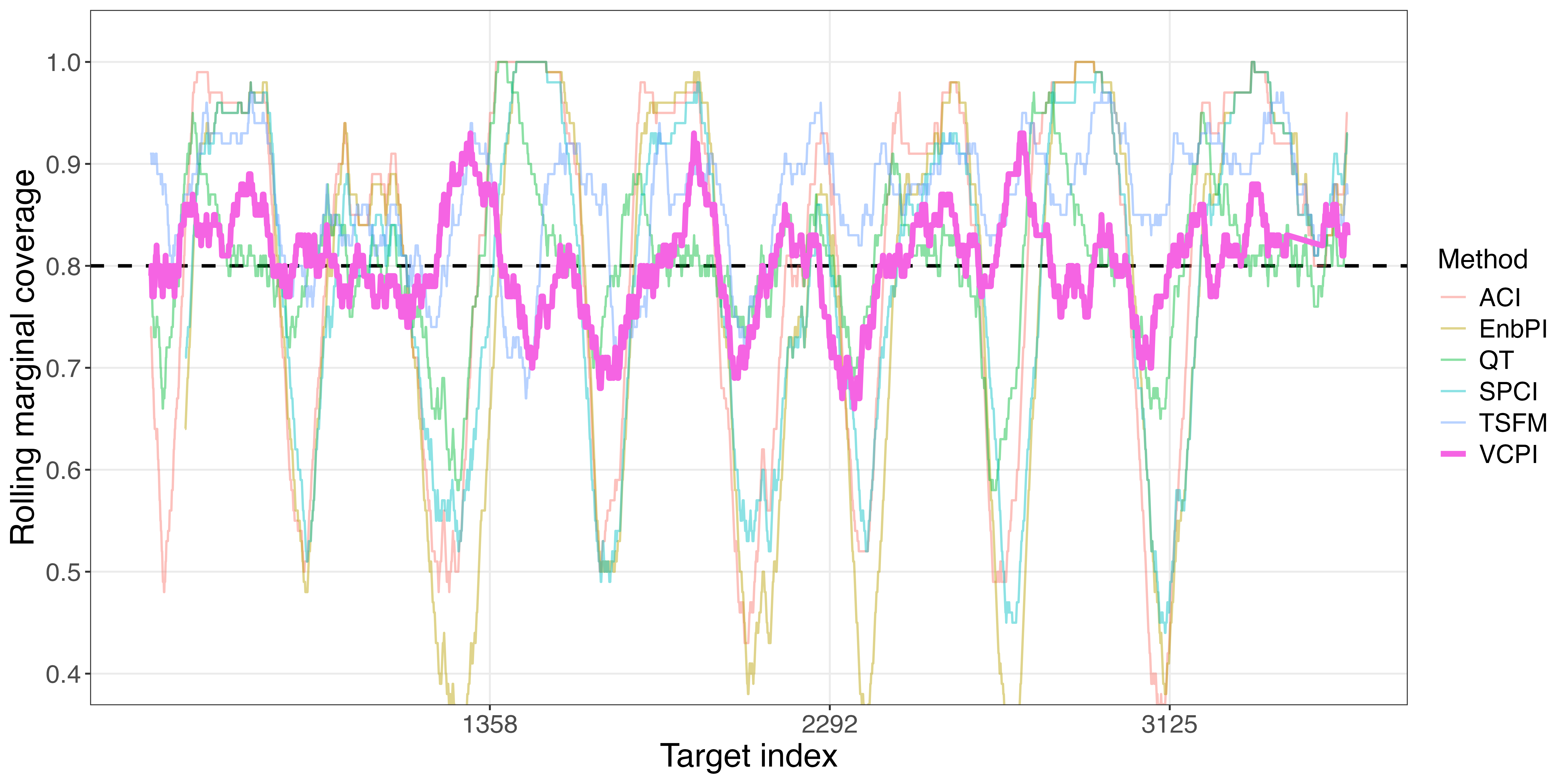}
\end{subfigure}
\begin{subfigure}{0.48\linewidth}
    \centering
    \includegraphics[width=\linewidth]{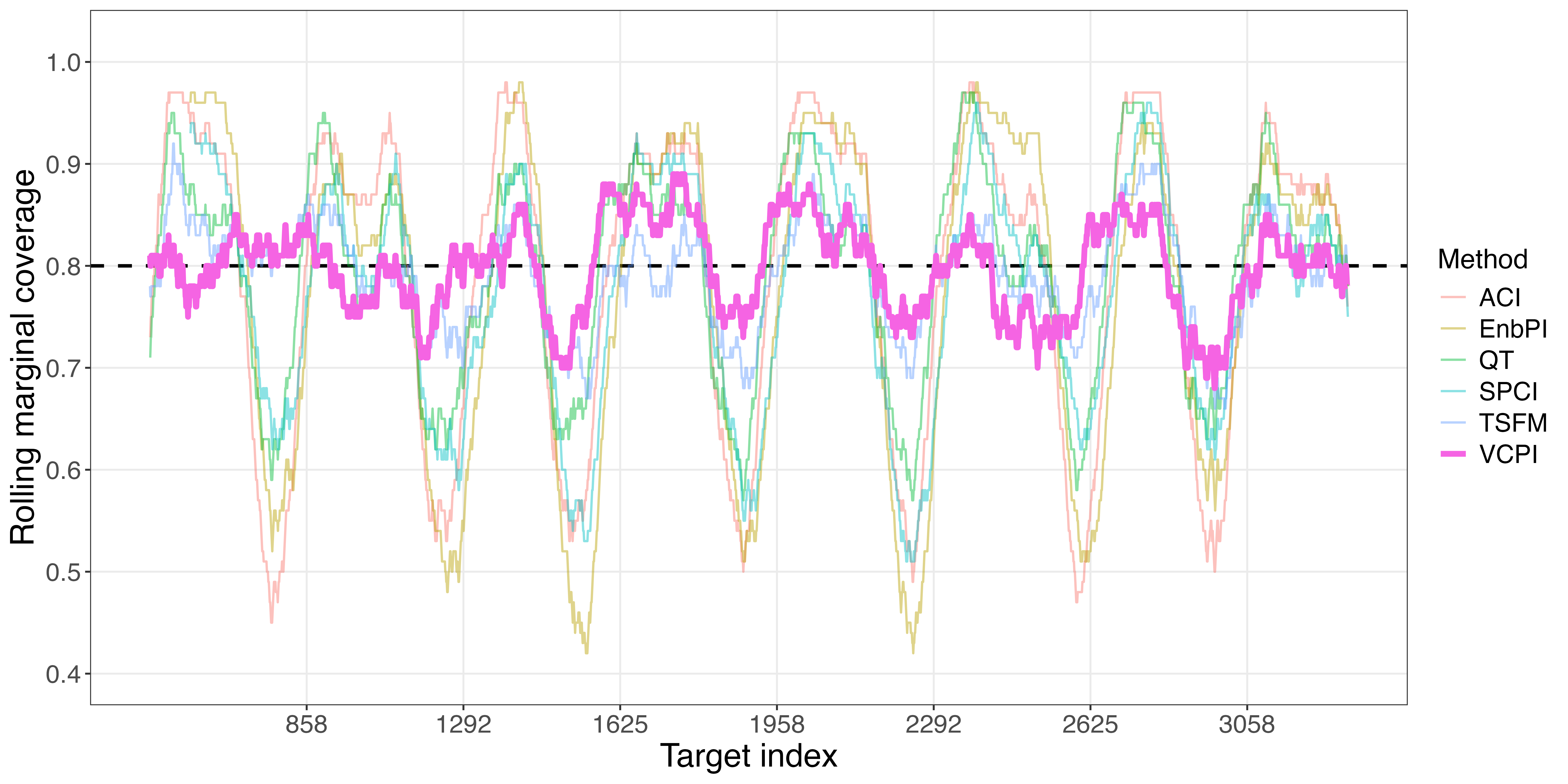}
\end{subfigure}
\hfill
\begin{subfigure}{0.48\linewidth}
    \centering
    \includegraphics[width=\linewidth]{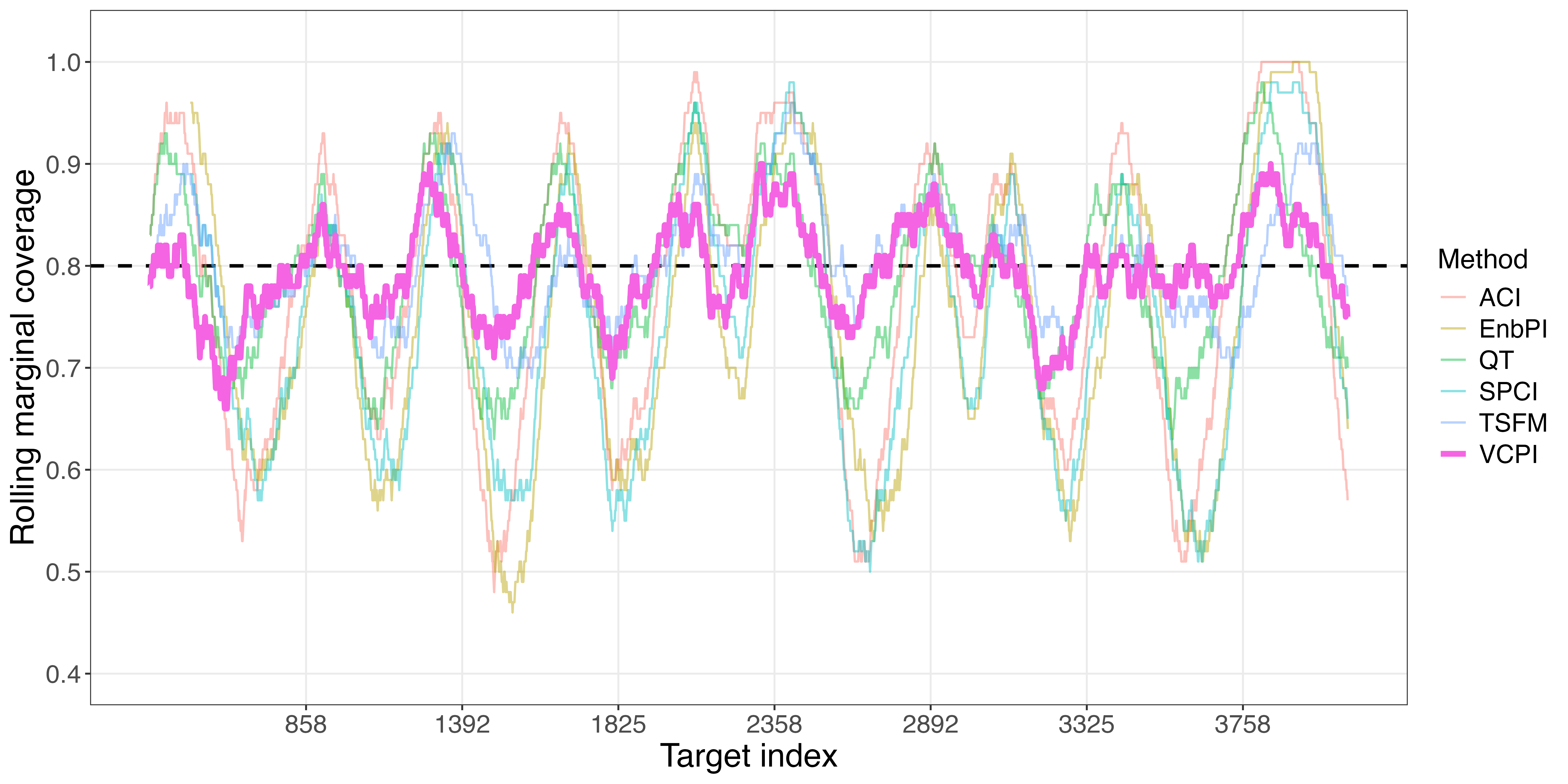}
\end{subfigure}
\caption{Rolling empirical coverage for the 80\% prediction intervals produced by TimesFM and the competing post-hoc uncertainty quantification methods across four daily time-series datasets: bike rental counts in the Washington metropolitan area via Capital Bikeshare~\citep{capital_bikeshare_data}, electricity demand in New York State from the New York Independent System Operator (NYISO)~\citep{nyis_electricity}, NO$_2$ concentration in Pasadena, CA (EPA AQS site 06-037-2005)~\citep{epa}, and streamflow of the Colorado River near Cisco, UT (USGS NWIS site 09180500, parameter 00060)~\citep{colorado}. Each panel reports rolling marginal coverage computed with a trailing window of \(k=100\) prediction instances for one-step-ahead forecasts (\(h=1\)). The dashed horizontal line indicates the nominal coverage level \(0.8\). These datasets exhibit pronounced time-varying volatility, so deviations from the dashed line indicate periods of local undercoverage or overcoverage. Curves closer to the dashed line therefore indicate better local calibration over time.}
\label{fig:result2}
\end{figure}
\newpage


\begin{landscape}
\begin{table}[p]\caption{\textbf{Comparison of 90\% Prediction Interval Performance Across Foundation Models and Calibration Methods.} Reported metrics are marginal coverage (MC), rolling root mean squared coverage error (RMSCE), and median interval width across datasets and forecast horizon blocks. Columns compare adaptive conformal inference (ACI), quantile tracking (QT), Ensemble Batch Prediction Interval (EnbPI), sequential predictive conformal inference (SPCI), and the proposed volatility-corrected prediction interval (VCPI), all targeting 90\% coverage (miscoverage level $\alpha = 0.1$). For ACI, the learning rate is fixed at 0.005, while for QT the learning rate multiplier is fixed at 0.05. RMSCE is the primary metric, as it measures local calibration under time-varying uncertainty. Blue bold entries indicate the smallest RMSCE within each model block. MC and Width are included  as complementary summaries and are not highlighted.  Native TSFM intervals are omitted because the TSFMs provide only decile-level quantiles, so native two-sided intervals are available at the 80\% level but not at the 90\% or 95\% levels.}\label{tab:cov_90}
\centering\begingroup\fontsize{9}{11}\selectfont

\resizebox{\ifdim\width>\linewidth\linewidth\else\width\fi}{!}{
\begin{tabular}[t]{lllccccccccccccccc}
\toprule
\multicolumn{3}{c}{ } & \multicolumn{5}{c}{Chronos} & \multicolumn{5}{c}{Moirai} & \multicolumn{5}{c}{TimesFM} \\
\cmidrule(l{3pt}r{3pt}){4-8} \cmidrule(l{3pt}r{3pt}){9-13} \cmidrule(l{3pt}r{3pt}){14-18}
Data & Metric & Horizon & ACI & QT & EnbPI & SPCI & VCPI & ACI & QT & EnbPI & SPCI & VCPI & ACI & QT & EnbPI & SPCI & VCPI\\
\midrule
 &  & 1 - 4 & 0.072 & 0.080 & 0.087 & 0.086 & \textbf{\textcolor{blue}{0.036}} & 0.076 & 0.084 & 0.092 & 0.091 & \textbf{\textcolor{blue}{0.036}} & 0.077 & 0.091 & 0.093 & 0.092 & \textbf{\textcolor{blue}{0.030}}\\

 &  & 5 - 8 & 0.070 & 0.077 & 0.083 & 0.086 & \textbf{\textcolor{blue}{0.043}} & 0.075 & 0.079 & 0.086 & 0.086 & \textbf{\textcolor{blue}{0.044}} & 0.078 & 0.092 & 0.092 & 0.088 & \textbf{\textcolor{blue}{0.046}}\\

 & \multirow{-3}{*}{\raggedright\arraybackslash RMSCE} & 9 - 16 & 0.081 & 0.082 & 0.097 & 0.099 & \textbf{\textcolor{blue}{0.054}} & 0.086 & 0.085 & 0.099 & 0.099 & \textbf{\textcolor{blue}{0.058}} & 0.084 & 0.091 & 0.096 & 0.093 & \textbf{\textcolor{blue}{0.049}}\\

 &  & 1 - 4 & 0.901 & 0.872 & 0.873 & 0.859 & 0.898 & 0.897 & 0.867 & 0.870 & 0.857 & 0.897 & 0.889 & 0.854 & 0.852 & 0.844 & 0.899\\

 &  & 5 - 8 & 0.902 & 0.873 & 0.870 & 0.850 & 0.899 & 0.895 & 0.867 & 0.862 & 0.849 & 0.896 & 0.893 & 0.856 & 0.849 & 0.841 & 0.892\\

 & \multirow{-3}{*}{\raggedright\arraybackslash MC} & 9 - 16 & 0.901 & 0.871 & 0.853 & 0.839 & 0.906 & 0.895 & 0.867 & 0.844 & 0.835 & 0.906 & 0.894 & 0.855 & 0.840 & 0.833 & 0.909\\

 &  & 1 - 4 & 3097.153 & 2627.731 & 2694.361 & 2518.061 & 2876.828 & 2902.080 & 2532.493 & 2590.459 & 2509.529 & 2557.974 & 3273.638 & 2775.616 & 2819.282 & 2704.788 & 3123.055\\

 &  & 5 - 8 & 3317.734 & 2756.687 & 2842.448 & 2578.464 & 3368.331 & 3051.825 & 2666.943 & 2674.346 & 2556.187 & 2897.297 & 3652.285 & 2972.430 & 2956.152 & 2880.105 & 3428.016\\

\multirow{-9}{*}{\raggedright\arraybackslash Bike Rental Counts} & \multirow{-3}{*}{\raggedright\arraybackslash Width} & 9 - 16 & 3642.114 & 3027.509 & 2978.339 & 2712.370 & 3770.102 & 3699.449 & 2893.981 & 2923.331 & 2820.353 & 3423.068 & 4335.819 & 3285.948 & 3117.253 & 3076.254 & 4095.263\\

\cmidrule{1-18}
 &  & 1 - 4 & 0.119 & 0.138 & 0.172 & 0.147 & \textbf{\textcolor{blue}{0.064}} & 0.121 & 0.144 & 0.194 & 0.165 & \textbf{\textcolor{blue}{0.055}} & 0.117 & 0.146 & 0.179 & 0.152 & \textbf{\textcolor{blue}{0.049}}\\

 &  & 5 - 8 & 0.129 & 0.142 & 0.210 & 0.210 & \textbf{\textcolor{blue}{0.093}} & 0.131 & 0.161 & 0.243 & 0.237 & \textbf{\textcolor{blue}{0.073}} & 0.131 & 0.156 & 0.223 & 0.211 & \textbf{\textcolor{blue}{0.082}}\\

 & \multirow{-3}{*}{\raggedright\arraybackslash RMSCE} & 9 - 16 & 0.158 & 0.154 & 0.253 & 0.278 & \textbf{\textcolor{blue}{0.116}} & 0.157 & 0.176 & 0.296 & 0.302 & \textbf{\textcolor{blue}{0.097}} & 0.161 & 0.165 & 0.265 & 0.273 & \textbf{\textcolor{blue}{0.110}}\\

 &  & 1 - 4 & 0.900 & 0.900 & 0.856 & 0.860 & 0.874 & 0.897 & 0.895 & 0.837 & 0.840 & 0.876 & 0.905 & 0.898 & 0.852 & 0.856 & 0.883\\

 &  & 5 - 8 & 0.898 & 0.904 & 0.833 & 0.813 & 0.851 & 0.895 & 0.897 & 0.806 & 0.791 & 0.863 & 0.904 & 0.903 & 0.828 & 0.818 & 0.865\\

 & \multirow{-3}{*}{\raggedright\arraybackslash MC} & 9 - 16 & 0.888 & 0.909 & 0.811 & 0.753 & 0.842 & 0.883 & 0.896 & 0.772 & 0.741 & 0.859 & 0.891 & 0.911 & 0.803 & 0.774 & 0.860\\

 &  & 1 - 4 & 3087.208 & 3057.038 & 2469.495 & 1925.863 & 1240.378 & 2782.374 & 3027.420 & 2178.848 & 1775.543 & 1102.066 & 3139.070 & 3039.353 & 2200.776 & 1989.554 & 1006.303\\

 &  & 5 - 8 & 6373.849 & 5738.037 & 4166.957 & 2770.612 & 2024.248 & 5527.801 & 6145.512 & 3538.764 & 2798.189 & 1989.427 & 6590.798 & 6223.138 & 4586.168 & 3010.536 & 1723.205\\

\multirow{-9}{*}{\raggedright\arraybackslash Colorado River} & \multirow{-3}{*}{\raggedright\arraybackslash Width} & 9 - 16 & 8100.949 & 8465.079 & 5659.709 & 3310.250 & 2844.784 & 9167.615 & 9538.315 & 4602.209 & 3258.030 & 2607.183 & 9527.121 & 9763.419 & 6581.834 & 3415.304 & 2300.358\\

\cmidrule{1-18}
 &  & 1 - 4 & 0.106 & 0.094 & 0.132 & 0.120 & \textbf{\textcolor{blue}{0.046}} & 0.107 & 0.092 & 0.132 & 0.114 & \textbf{\textcolor{blue}{0.049}} & 0.115 & 0.096 & 0.139 & 0.124 & \textbf{\textcolor{blue}{0.044}}\\

 &  & 5 - 8 & 0.117 & 0.099 & 0.155 & 0.137 & \textbf{\textcolor{blue}{0.058}} & 0.118 & 0.101 & 0.157 & 0.136 & \textbf{\textcolor{blue}{0.064}} & 0.122 & 0.107 & 0.158 & 0.145 & \textbf{\textcolor{blue}{0.050}}\\

 & \multirow{-3}{*}{\raggedright\arraybackslash RMSCE} & 9 - 16 & 0.124 & 0.104 & 0.177 & 0.159 & \textbf{\textcolor{blue}{0.077}} & 0.130 & 0.110 & 0.178 & 0.155 & \textbf{\textcolor{blue}{0.076}} & 0.132 & 0.119 & 0.184 & 0.180 & \textbf{\textcolor{blue}{0.054}}\\

 &  & 1 - 4 & 0.896 & 0.901 & 0.869 & 0.861 & 0.888 & 0.900 & 0.906 & 0.876 & 0.871 & 0.883 & 0.892 & 0.897 & 0.861 & 0.851 & 0.883\\

 &  & 5 - 8 & 0.893 & 0.899 & 0.856 & 0.846 & 0.878 & 0.898 & 0.905 & 0.860 & 0.850 & 0.883 & 0.890 & 0.892 & 0.842 & 0.827 & 0.877\\

 & \multirow{-3}{*}{\raggedright\arraybackslash MC} & 9 - 16 & 0.890 & 0.895 & 0.836 & 0.813 & 0.867 & 0.893 & 0.901 & 0.839 & 0.824 & 0.875 & 0.887 & 0.891 & 0.821 & 0.793 & 0.883\\

 &  & 1 - 4 & 116384.078 & 110195.446 & 107055.712 & 98077.359 & 81084.163 & 115000.761 & 109038.863 & 108638.016 & 98699.759 & 81676.942 & 116846.183 & 106857.567 & 106727.970 & 93959.005 & 75753.027\\

 &  & 5 - 8 & 141531.749 & 130939.373 & 124067.361 & 109114.399 & 97672.851 & 145670.765 & 132414.240 & 127504.163 & 108670.833 & 96167.353 & 143474.428 & 125091.692 & 119069.932 & 103583.796 & 91486.627\\

\multirow{-9}{*}{\raggedright\arraybackslash Electricity Demand} & \multirow{-3}{*}{\raggedright\arraybackslash Width} & 9 - 16 & 150448.712 & 148578.762 & 134686.886 & 113963.306 & 113958.115 & 163369.915 & 147700.249 & 138092.281 & 113798.241 & 108440.264 & 154707.546 & 139415.672 & 128287.408 & 104061.149 & 102798.667\\

\cmidrule{1-18}
 &  & 1 - 4 & 0.098 & 0.079 & 0.137 & 0.134 & \textbf{\textcolor{blue}{0.050}} & 0.095 & 0.075 & 0.134 & 0.131 & \textbf{\textcolor{blue}{0.045}} & 0.102 & 0.082 & 0.148 & 0.155 & \textbf{\textcolor{blue}{0.048}}\\

 &  & 5 - 8 & 0.104 & 0.087 & 0.150 & 0.159 & \textbf{\textcolor{blue}{0.064}} & 0.103 & 0.082 & 0.150 & 0.153 & \textbf{\textcolor{blue}{0.056}} & 0.106 & 0.084 & 0.168 & 0.180 & \textbf{\textcolor{blue}{0.062}}\\

 & \multirow{-3}{*}{\raggedright\arraybackslash RMSCE} & 9 - 16 & 0.110 & 0.091 & 0.164 & 0.171 & \textbf{\textcolor{blue}{0.076}} & 0.110 & 0.087 & 0.162 & 0.169 & \textbf{\textcolor{blue}{0.067}} & 0.113 & 0.089 & 0.188 & 0.194 & \textbf{\textcolor{blue}{0.072}}\\

 &  & 1 - 4 & 0.895 & 0.902 & 0.856 & 0.844 & 0.884 & 0.894 & 0.899 & 0.849 & 0.840 & 0.892 & 0.883 & 0.889 & 0.833 & 0.818 & 0.885\\

 &  & 5 - 8 & 0.893 & 0.900 & 0.845 & 0.824 & 0.885 & 0.893 & 0.897 & 0.840 & 0.821 & 0.891 & 0.884 & 0.889 & 0.819 & 0.797 & 0.889\\

 & \multirow{-3}{*}{\raggedright\arraybackslash MC} & 9 - 16 & 0.891 & 0.898 & 0.832 & 0.814 & 0.878 & 0.893 & 0.897 & 0.828 & 0.809 & 0.889 & 0.880 & 0.886 & 0.805 & 0.782 & 0.883\\

 &  & 1 - 4 & 20.923 & 19.387 & 18.422 & 17.364 & 16.395 & 20.583 & 19.683 & 18.277 & 17.324 & 17.505 & 19.994 & 19.552 & 17.703 & 16.897 & 17.862\\

 &  & 5 - 8 & 21.973 & 20.681 & 19.197 & 17.877 & 17.212 & 21.549 & 20.495 & 19.048 & 18.005 & 18.274 & 21.573 & 21.000 & 18.870 & 17.796 & 19.377\\

\multirow{-9}{*}{\raggedright\arraybackslash NO2 Concentration} & \multirow{-3}{*}{\raggedright\arraybackslash Width} & 9 - 16 & 22.536 & 21.156 & 19.619 & 18.123 & 17.591 & 22.244 & 20.941 & 19.475 & 18.316 & 18.524 & 22.386 & 22.138 & 19.053 & 17.987 & 19.933\\
\bottomrule
\end{tabular}}
\endgroup{}
\end{table}
\end{landscape}

\newpage

\begin{landscape}
\begin{table}[p]\caption{\textbf{Comparison of 95\% Prediction Interval Performance Across Foundation Models and Calibration Methods.} Reported metrics are marginal coverage (MC), rolling root mean squared coverage error (RMSCE), and median interval width across datasets and forecast horizon blocks. Columns compare adaptive conformal inference (ACI), quantile tracking (QT), Ensemble Batch Prediction Interval (EnbPI), sequential predictive conformal inference (SPCI), and the proposed volatility-corrected prediction interval (VCPI), all targeting 95\% coverage (miscoverage level $\alpha = 0.05$). For ACI, the learning rate is fixed at 0.005, while for QT the learning rate multiplier is fixed at 0.05. RMSCE is the primary metric, as it measures local calibration under time-varying uncertainty. Blue bold entries indicate the smallest RMSCE within each model block. MC and Width are included  as complementary summaries and are not highlighted. Native TSFM intervals are omitted because the TSFMs provide only decile-level quantiles, so native two-sided intervals are available at the 80\% level but not at the 90\% or 95\% levels. }\label{tab:cov_95}
\centering\begingroup\fontsize{9}{11}\selectfont

\resizebox{\ifdim\width>\linewidth\linewidth\else\width\fi}{!}{
\begin{tabular}[t]{lllccccccccccccccc}
\toprule
\multicolumn{3}{c}{ } & \multicolumn{5}{c}{Chronos} & \multicolumn{5}{c}{Moirai} & \multicolumn{5}{c}{TimesFM} \\
\cmidrule(l{3pt}r{3pt}){4-8} \cmidrule(l{3pt}r{3pt}){9-13} \cmidrule(l{3pt}r{3pt}){14-18}
Data & Metric & Horizon & ACI & QT & EnbPI & SPCI & VCPI & ACI & QT & EnbPI & SPCI & VCPI & ACI & QT & EnbPI & SPCI & VCPI\\
\midrule
 &  & 1 - 4 & 0.048 & 0.066 & 0.060 & 0.071 & \textbf{\textcolor{blue}{0.031}} & 0.049 & 0.071 & 0.062 & 0.073 & \textbf{\textcolor{blue}{0.030}} & 0.049 & 0.079 & 0.065 & 0.081 & \textbf{\textcolor{blue}{0.026}}\\

 &  & 5 - 8 & 0.050 & 0.067 & 0.062 & 0.072 & \textbf{\textcolor{blue}{0.039}} & 0.054 & 0.065 & 0.061 & 0.077 & \textbf{\textcolor{blue}{0.042}} & 0.054 & 0.078 & 0.067 & 0.076 & \textbf{\textcolor{blue}{0.040}}\\

 & \multirow{-3}{*}{\raggedright\arraybackslash RMSCE} & 9 - 16 & 0.057 & 0.070 & 0.069 & 0.089 & \textbf{\textcolor{blue}{0.046}} & 0.064 & 0.071 & 0.071 & 0.090 & \textbf{\textcolor{blue}{0.051}} & 0.062 & 0.077 & 0.072 & 0.083 & \textbf{\textcolor{blue}{0.036}}\\

 &  & 1 - 4 & 0.951 & 0.918 & 0.927 & 0.910 & 0.935 & 0.947 & 0.915 & 0.923 & 0.909 & 0.934 & 0.941 & 0.901 & 0.909 & 0.893 & 0.934\\

 &  & 5 - 8 & 0.951 & 0.917 & 0.926 & 0.908 & 0.935 & 0.947 & 0.914 & 0.921 & 0.900 & 0.933 & 0.944 & 0.901 & 0.910 & 0.895 & 0.930\\

 & \multirow{-3}{*}{\raggedright\arraybackslash MC} & 9 - 16 & 0.950 & 0.917 & 0.918 & 0.892 & 0.939 & 0.946 & 0.916 & 0.911 & 0.887 & 0.937 & 0.946 & 0.904 & 0.905 & 0.886 & 0.941\\

 &  & 1 - 4 & 4773.527 & 3318.464 & 3940.814 & 3252.162 & 3789.940 & 4820.377 & 3286.945 & 3882.936 & 3332.718 & 3322.660 & 5669.815 & 3579.702 & 4128.847 & 3526.129 & 4037.677\\

 &  & 5 - 8 & 5490.920 & 3537.164 & 4239.257 & 3378.212 & 4467.621 & 5489.345 & 3443.518 & 4095.534 & 3411.778 & 3744.569 & 6125.017 & 3730.610 & 4352.479 & 3699.655 & 4401.244\\

\multirow{-9}{*}{\raggedright\arraybackslash Bike Rental Counts} & \multirow{-3}{*}{\raggedright\arraybackslash Width} & 9 - 16 & 5910.358 & 3759.153 & 4169.630 & 3499.592 & 4902.275 & 6160.995 & 3726.812 & 4169.204 & 3642.943 & 4280.696 & 6964.580 & 4144.998 & 4382.243 & 3978.838 & 5291.024\\

\cmidrule{1-18}
 &  & 1 - 4 & 0.067 & 0.102 & 0.138 & 0.133 & \textbf{\textcolor{blue}{0.062}} & 0.071 & 0.104 & 0.151 & 0.148 & \textbf{\textcolor{blue}{0.048}} & 0.068 & 0.107 & 0.138 & 0.133 & \textbf{\textcolor{blue}{0.046}}\\

 &  & 5 - 8 & \textbf{\textcolor{blue}{0.088}} & 0.101 & 0.186 & 0.195 & 0.095 & 0.090 & 0.114 & 0.206 & 0.226 & \textbf{\textcolor{blue}{0.075}} & 0.089 & 0.111 & 0.196 & 0.196 & \textbf{\textcolor{blue}{0.081}}\\

 & \multirow{-3}{*}{\raggedright\arraybackslash RMSCE} & 9 - 16 & 0.123 & \textbf{\textcolor{blue}{0.112}} & 0.238 & 0.272 & 0.113 & 0.129 & 0.125 & 0.272 & 0.303 & \textbf{\textcolor{blue}{0.095}} & 0.134 & 0.121 & 0.246 & 0.269 & \textbf{\textcolor{blue}{0.108}}\\

 &  & 1 - 4 & 0.946 & 0.942 & 0.897 & 0.892 & 0.912 & 0.944 & 0.938 & 0.887 & 0.879 & 0.919 & 0.949 & 0.941 & 0.898 & 0.892 & 0.921\\

 &  & 5 - 8 & 0.938 & 0.947 & 0.870 & 0.852 & 0.885 & 0.935 & 0.939 & 0.856 & 0.826 & 0.896 & 0.941 & 0.946 & 0.866 & 0.855 & 0.898\\

 & \multirow{-3}{*}{\raggedright\arraybackslash MC} & 9 - 16 & 0.922 & 0.953 & 0.843 & 0.792 & 0.879 & 0.914 & 0.940 & 0.815 & 0.772 & 0.891 & 0.920 & 0.950 & 0.838 & 0.804 & 0.890\\

 &  & 1 - 4 & 5221.400 & 4709.039 & 4044.923 & 2861.108 & 1521.479 & 4895.079 & 4659.019 & 3578.218 & 2655.018 & 1357.410 & 5351.880 & 4775.165 & 3696.155 & 2891.804 & 1243.813\\

 &  & 5 - 8 & 9860.251 & 9185.780 & 6984.862 & 4533.282 & 2450.833 & 10456.173 & 9639.514 & 6001.021 & 4274.660 & 2386.617 & 11106.151 & 10107.418 & 6791.348 & 4691.584 & 2060.211\\

\multirow{-9}{*}{\raggedright\arraybackslash Colorado River} & \multirow{-3}{*}{\raggedright\arraybackslash Width} & 9 - 16 & 12238.822 & 13734.451 & 8567.210 & 4710.850 & 3484.294 & 12926.449 & 13974.072 & 7330.203 & 4480.864 & 3125.620 & 13995.405 & 15871.820 & 9349.942 & 4628.788 & 2742.443\\

\cmidrule{1-18}
 &  & 1 - 4 & 0.065 & 0.058 & 0.097 & 0.105 & \textbf{\textcolor{blue}{0.038}} & 0.062 & 0.055 & 0.098 & 0.097 & \textbf{\textcolor{blue}{0.042}} & 0.067 & 0.061 & 0.101 & 0.107 & \textbf{\textcolor{blue}{0.038}}\\

 &  & 5 - 8 & 0.070 & \textbf{\textcolor{blue}{0.059}} & 0.112 & 0.124 & 0.060 & 0.069 & \textbf{\textcolor{blue}{0.057}} & 0.121 & 0.119 & 0.058 & 0.071 & 0.063 & 0.124 & 0.132 & \textbf{\textcolor{blue}{0.048}}\\

 & \multirow{-3}{*}{\raggedright\arraybackslash RMSCE} & 9 - 16 & 0.085 & \textbf{\textcolor{blue}{0.065}} & 0.145 & 0.148 & 0.074 & 0.085 & \textbf{\textcolor{blue}{0.063}} & 0.142 & 0.142 & 0.067 & 0.094 & 0.075 & 0.157 & 0.168 & \textbf{\textcolor{blue}{0.051}}\\

 &  & 1 - 4 & 0.947 & 0.953 & 0.918 & 0.904 & 0.934 & 0.948 & 0.957 & 0.919 & 0.911 & 0.930 & 0.945 & 0.949 & 0.910 & 0.901 & 0.931\\

 &  & 5 - 8 & 0.946 & 0.952 & 0.905 & 0.888 & 0.918 & 0.946 & 0.956 & 0.903 & 0.894 & 0.925 & 0.941 & 0.946 & 0.888 & 0.874 & 0.924\\

 & \multirow{-3}{*}{\raggedright\arraybackslash MC} & 9 - 16 & 0.938 & 0.948 & 0.882 & 0.862 & 0.905 & 0.936 & 0.952 & 0.885 & 0.869 & 0.918 & 0.927 & 0.943 & 0.867 & 0.844 & 0.922\\

 &  & 1 - 4 & 157792.453 & 153375.982 & 137724.888 & 122674.994 & 97714.711 & 162701.506 & 154596.564 & 140013.863 & 123549.219 & 98462.084 & 152002.265 & 142279.330 & 134809.249 & 117063.688 & 90940.057\\

 &  & 5 - 8 & 189680.613 & 177228.251 & 161788.849 & 135482.898 & 116907.255 & 193368.982 & 179900.592 & 163167.592 & 133304.234 & 115328.265 & 182205.548 & 165653.428 & 151286.762 & 128324.077 & 109564.571\\

\multirow{-9}{*}{\raggedright\arraybackslash Electricity Demand} & \multirow{-3}{*}{\raggedright\arraybackslash Width} & 9 - 16 & 204626.726 & 189447.993 & 167181.128 & 140758.591 & 135703.307 & 218972.678 & 200342.679 & 173755.607 & 139656.662 & 129545.226 & 203586.072 & 185601.910 & 156439.662 & 127393.767 & 122982.831\\

\cmidrule{1-18}
 &  & 1 - 4 & 0.056 & 0.050 & 0.097 & 0.111 & \textbf{\textcolor{blue}{0.036}} & 0.057 & 0.046 & 0.098 & 0.109 & \textbf{\textcolor{blue}{0.031}} & 0.060 & 0.053 & 0.110 & 0.130 & \textbf{\textcolor{blue}{0.036}}\\

 &  & 5 - 8 & 0.061 & 0.054 & 0.114 & 0.137 & \textbf{\textcolor{blue}{0.049}} & 0.060 & 0.051 & 0.111 & 0.130 & \textbf{\textcolor{blue}{0.038}} & 0.061 & 0.055 & 0.127 & 0.157 & \textbf{\textcolor{blue}{0.050}}\\

 & \multirow{-3}{*}{\raggedright\arraybackslash RMSCE} & 9 - 16 & 0.065 & \textbf{\textcolor{blue}{0.054}} & 0.127 & 0.152 & 0.056 & 0.064 & 0.053 & 0.121 & 0.145 & \textbf{\textcolor{blue}{0.046}} & 0.067 & \textbf{\textcolor{blue}{0.053}} & 0.141 & 0.170 & 0.056\\

 &  & 1 - 4 & 0.947 & 0.954 & 0.911 & 0.893 & 0.934 & 0.947 & 0.951 & 0.907 & 0.891 & 0.942 & 0.939 & 0.944 & 0.897 & 0.874 & 0.938\\

 &  & 5 - 8 & 0.945 & 0.953 & 0.898 & 0.872 & 0.933 & 0.946 & 0.950 & 0.896 & 0.874 & 0.940 & 0.939 & 0.943 & 0.883 & 0.853 & 0.938\\

 & \multirow{-3}{*}{\raggedright\arraybackslash MC} & 9 - 16 & 0.943 & 0.950 & 0.887 & 0.860 & 0.930 & 0.946 & 0.949 & 0.888 & 0.864 & 0.938 & 0.937 & 0.943 & 0.874 & 0.841 & 0.933\\

 &  & 1 - 4 & 25.702 & 24.438 & 22.460 & 20.596 & 19.641 & 25.304 & 24.046 & 22.383 & 20.657 & 20.901 & 24.008 & 23.721 & 21.283 & 19.978 & 21.308\\

 &  & 5 - 8 & 27.279 & 25.760 & 23.272 & 21.132 & 20.556 & 26.213 & 25.248 & 22.909 & 21.242 & 21.824 & 25.634 & 25.246 & 22.630 & 20.830 & 23.126\\

\multirow{-9}{*}{\raggedright\arraybackslash NO2 Concentration} & \multirow{-3}{*}{\raggedright\arraybackslash Width} & 9 - 16 & 28.076 & 26.176 & 23.687 & 21.524 & 21.009 & 27.187 & 25.692 & 23.424 & 21.546 & 22.130 & 27.502 & 26.136 & 23.106 & 21.095 & 23.777\\
\bottomrule
\end{tabular}}
\endgroup{}
\end{table}
\end{landscape}

\newpage

\begin{table}[!htbp]\caption{\textbf{ACI Sensitivity at the 80\% Prediction Level.}Performance of adaptive conformal inference (ACI) at the 80\% prediction level across datasets, forecast horizon blocks, and foundation models, shown over a grid of learning-rate values \(\gamma \in (0.001,0.005,0.01)\). Reported metrics are marginal coverage (MC), rolling root mean squared coverage error (RMSCE), and median interval width. Columns are grouped by foundation model, with learning-rate values shown within each model block.}\label{tab:aci_sensitivity}
\centering\begingroup\fontsize{9}{11}\selectfont

\resizebox{\ifdim\width>\linewidth\linewidth\else\width\fi}{!}{
\begin{tabular}[t]{lllccccccccc}
\toprule
\multicolumn{3}{c}{ } & \multicolumn{3}{c}{Chronos} & \multicolumn{3}{c}{Moirai} & \multicolumn{3}{c}{TimesFM} \\
\cmidrule(l{3pt}r{3pt}){4-6} \cmidrule(l{3pt}r{3pt}){7-9} \cmidrule(l{3pt}r{3pt}){10-12}
Data & Metric & Horizon & 0.001 & 0.005 & 0.01 & 0.001 & 0.005 & 0.01 & 0.001 & 0.005 & 0.01\\
\midrule
 &  & 1 - 4 & 0.105 & 0.100 & 0.093 & 0.112 & 0.105 & 0.095 & 0.112 & 0.108 & 0.097\\

 &  & 5 - 8 & 0.096 & 0.092 & 0.084 & 0.106 & 0.099 & 0.093 & 0.108 & 0.101 & 0.093\\

 & \multirow{-3}{*}{\raggedright\arraybackslash RMSCE} & 9 - 16 & 0.109 & 0.106 & 0.100 & 0.123 & 0.115 & 0.107 & 0.116 & 0.111 & 0.105\\

 &  & 1 - 4 & 0.795 & 0.803 & 0.805 & 0.789 & 0.799 & 0.802 & 0.775 & 0.788 & 0.792\\

 &  & 5 - 8 & 0.796 & 0.803 & 0.806 & 0.789 & 0.799 & 0.800 & 0.776 & 0.788 & 0.792\\

 & \multirow{-3}{*}{\raggedright\arraybackslash MC} & 9 - 16 & 0.793 & 0.803 & 0.805 & 0.786 & 0.799 & 0.802 & 0.776 & 0.792 & 0.798\\

 &  & 1 - 4 & 1913.861 & 1929.987 & 1948.136 & 1821.386 & 1831.740 & 1868.208 & 1902.288 & 1952.828 & 2048.574\\

 &  & 5 - 8 & 1944.720 & 2045.242 & 2007.847 & 1907.560 & 1974.620 & 2014.536 & 2102.228 & 2150.538 & 2237.407\\

\multirow{-9}{*}{\raggedright\arraybackslash Bike Rental Counts} & \multirow{-3}{*}{\raggedright\arraybackslash Width} & 9 - 16 & 2151.040 & 2222.898 & 2233.597 & 2049.777 & 2136.972 & 2205.839 & 2391.697 & 2593.445 & 2613.166\\

\cmidrule{1-12}
 &  & 1 - 4 & 0.204 & 0.192 & 0.170 & 0.212 & 0.201 & 0.184 & 0.208 & 0.192 & 0.175\\

 &  & 5 - 8 & 0.220 & 0.206 & 0.192 & 0.231 & 0.216 & 0.209 & 0.224 & 0.205 & 0.197\\

 & \multirow{-3}{*}{\raggedright\arraybackslash RMSCE} & 9 - 16 & 0.229 & 0.210 & 0.194 & 0.252 & 0.236 & 0.233 & 0.237 & 0.222 & 0.221\\

 &  & 1 - 4 & 0.801 & 0.804 & 0.803 & 0.791 & 0.798 & 0.798 & 0.802 & 0.811 & 0.810\\

 &  & 5 - 8 & 0.797 & 0.803 & 0.802 & 0.787 & 0.797 & 0.796 & 0.801 & 0.813 & 0.812\\

 & \multirow{-3}{*}{\raggedright\arraybackslash MC} & 9 - 16 & 0.798 & 0.803 & 0.794 & 0.786 & 0.797 & 0.794 & 0.803 & 0.812 & 0.807\\

 &  & 1 - 4 & 1332.573 & 1226.021 & 1159.153 & 1249.351 & 1201.905 & 1091.953 & 1215.881 & 1204.575 & 1143.343\\

 &  & 5 - 8 & 2270.000 & 2263.809 & 2198.704 & 2332.924 & 2157.600 & 2082.610 & 2188.148 & 2066.674 & 2031.951\\

\multirow{-9}{*}{\raggedright\arraybackslash Colorado River} & \multirow{-3}{*}{\raggedright\arraybackslash Width} & 9 - 16 & 3320.400 & 3320.557 & 3132.160 & 3659.320 & 3088.864 & 2705.087 & 3463.434 & 2591.209 & 2443.542\\

\cmidrule{1-12}
 &  & 1 - 4 & 0.163 & 0.154 & 0.138 & 0.161 & 0.151 & 0.133 & 0.170 & 0.161 & 0.139\\

 &  & 5 - 8 & 0.175 & 0.171 & 0.156 & 0.185 & 0.174 & 0.154 & 0.183 & 0.172 & 0.150\\

 & \multirow{-3}{*}{\raggedright\arraybackslash RMSCE} & 9 - 16 & 0.176 & 0.173 & 0.157 & 0.188 & 0.177 & 0.156 & 0.188 & 0.180 & 0.158\\

 &  & 1 - 4 & 0.796 & 0.795 & 0.796 & 0.805 & 0.802 & 0.798 & 0.788 & 0.789 & 0.790\\

 &  & 5 - 8 & 0.789 & 0.791 & 0.792 & 0.800 & 0.801 & 0.798 & 0.787 & 0.787 & 0.787\\

 & \multirow{-3}{*}{\raggedright\arraybackslash MC} & 9 - 16 & 0.787 & 0.788 & 0.790 & 0.800 & 0.799 & 0.795 & 0.786 & 0.787 & 0.787\\

 &  & 1 - 4 & 76154.504 & 76404.939 & 72882.385 & 77090.168 & 76552.150 & 70876.746 & 74057.299 & 73913.372 & 71963.344\\

 &  & 5 - 8 & 93091.915 & 93167.307 & 88684.072 & 98341.544 & 96834.728 & 89269.801 & 86684.421 & 86414.134 & 82579.533\\

\multirow{-9}{*}{\raggedright\arraybackslash Electricity Demand} & \multirow{-3}{*}{\raggedright\arraybackslash Width} & 9 - 16 & 105815.419 & 106781.678 & 102616.082 & 105163.868 & 103152.068 & 95164.541 & 91758.015 & 88704.532 & 85580.901\\

\cmidrule{1-12}
 &  & 1 - 4 & 0.164 & 0.157 & 0.147 & 0.157 & 0.152 & 0.143 & 0.167 & 0.160 & 0.148\\

 &  & 5 - 8 & 0.173 & 0.168 & 0.157 & 0.167 & 0.160 & 0.151 & 0.175 & 0.170 & 0.161\\

 & \multirow{-3}{*}{\raggedright\arraybackslash RMSCE} & 9 - 16 & 0.178 & 0.174 & 0.165 & 0.174 & 0.170 & 0.163 & 0.180 & 0.178 & 0.171\\

 &  & 1 - 4 & 0.792 & 0.795 & 0.794 & 0.782 & 0.790 & 0.794 & 0.765 & 0.773 & 0.781\\

 &  & 5 - 8 & 0.788 & 0.792 & 0.791 & 0.780 & 0.791 & 0.794 & 0.762 & 0.771 & 0.780\\

 & \multirow{-3}{*}{\raggedright\arraybackslash MC} & 9 - 16 & 0.784 & 0.788 & 0.788 & 0.779 & 0.789 & 0.796 & 0.756 & 0.766 & 0.777\\

 &  & 1 - 4 & 14.705 & 14.549 & 14.080 & 14.705 & 14.892 & 14.478 & 14.085 & 14.556 & 14.542\\

 &  & 5 - 8 & 15.598 & 15.482 & 14.452 & 15.844 & 15.832 & 15.301 & 15.286 & 15.603 & 15.976\\

\multirow{-9}{*}{\raggedright\arraybackslash NO2 Concentration} & \multirow{-3}{*}{\raggedright\arraybackslash Width} & 9 - 16 & 15.569 & 15.585 & 14.670 & 15.939 & 16.332 & 15.860 & 15.515 & 16.000 & 17.039\\
\bottomrule
\end{tabular}}
\endgroup{}
\end{table}

\newpage

\begin{table}[!htbp]\caption{\textbf{QT Sensitivity at the 80\% Prediction Level.} 
Performance of quantile tracking (QT) at the 80\% prediction level across datasets, forecast horizon blocks, and foundation models. Reported metrics are marginal coverage (MC), rolling root mean squared coverage error (RMSCE), and median interval width. Columns are grouped by foundation model. Within each model block, results are shown over a grid of step-size multipliers \(\gamma \in \{0.01,0.05,0.1\}\), where the step size at each forecast origin is $\gamma \times \overline{|R|}$, where $\overline{|R|}$ denotes the mean absolute residual computed over the calibration window.}\label{tab:qt_sensitivity}
\centering\begingroup\fontsize{9}{11}\selectfont

\resizebox{\ifdim\width>\linewidth\linewidth\else\width\fi}{!}{
\begin{tabular}[t]{lllccccccccc}
\toprule
\multicolumn{3}{c}{ } & \multicolumn{3}{c}{Chronos} & \multicolumn{3}{c}{Moirai} & \multicolumn{3}{c}{TimesFM} \\
\cmidrule(l{3pt}r{3pt}){4-6} \cmidrule(l{3pt}r{3pt}){7-9} \cmidrule(l{3pt}r{3pt}){10-12}
Data & Metric & Horizon & 0.01 & 0.05 & 0.1 & 0.01 & 0.05 & 0.1 & 0.01 & 0.05 & 0.1\\
\midrule
 &  & 1 - 4 & 0.131 & 0.096 & 0.085 & 0.138 & 0.102 & 0.087 & 0.151 & 0.105 & 0.089\\

 &  & 5 - 8 & 0.124 & 0.087 & 0.079 & 0.131 & 0.095 & 0.081 & 0.144 & 0.099 & 0.085\\

 & \multirow{-3}{*}{\raggedright\arraybackslash RMSCE} & 9 - 16 & 0.131 & 0.099 & 0.092 & 0.142 & 0.106 & 0.091 & 0.156 & 0.107 & 0.094\\

 &  & 1 - 4 & 0.731 & 0.782 & 0.792 & 0.726 & 0.777 & 0.789 & 0.701 & 0.763 & 0.779\\

 &  & 5 - 8 & 0.730 & 0.782 & 0.793 & 0.724 & 0.776 & 0.788 & 0.708 & 0.766 & 0.780\\

 & \multirow{-3}{*}{\raggedright\arraybackslash MC} & 9 - 16 & 0.735 & 0.782 & 0.792 & 0.725 & 0.777 & 0.790 & 0.705 & 0.768 & 0.788\\

 &  & 1 - 4 & 1563.257 & 1802.969 & 1910.362 & 1449.619 & 1756.069 & 1865.441 & 1620.849 & 1916.520 & 2082.750\\

 &  & 5 - 8 & 1601.792 & 1858.352 & 1984.672 & 1506.959 & 1846.830 & 1929.636 & 1712.053 & 2035.227 & 2144.623\\

\multirow{-9}{*}{\raggedright\arraybackslash Bike Rental Counts} & \multirow{-3}{*}{\raggedright\arraybackslash Width} & 9 - 16 & 1779.937 & 2076.541 & 2238.096 & 1687.893 & 2036.814 & 2236.055 & 1967.631 & 2361.087 & 2537.229\\

\cmidrule{1-12}
 &  & 1 - 4 & 0.200 & 0.177 & 0.149 & 0.209 & 0.190 & 0.160 & 0.210 & 0.187 & 0.155\\

 &  & 5 - 8 & 0.209 & 0.187 & 0.153 & 0.233 & 0.205 & 0.173 & 0.227 & 0.197 & 0.167\\

 & \multirow{-3}{*}{\raggedright\arraybackslash RMSCE} & 9 - 16 & 0.213 & 0.188 & 0.147 & 0.251 & 0.223 & 0.185 & 0.240 & 0.209 & 0.174\\

 &  & 1 - 4 & 0.806 & 0.805 & 0.803 & 0.800 & 0.800 & 0.799 & 0.804 & 0.810 & 0.807\\

 &  & 5 - 8 & 0.816 & 0.808 & 0.804 & 0.799 & 0.800 & 0.799 & 0.808 & 0.813 & 0.808\\

 & \multirow{-3}{*}{\raggedright\arraybackslash MC} & 9 - 16 & 0.821 & 0.807 & 0.804 & 0.800 & 0.802 & 0.799 & 0.819 & 0.815 & 0.810\\

 &  & 1 - 4 & 1456.868 & 1510.577 & 1466.340 & 1416.112 & 1440.676 & 1460.942 & 1360.871 & 1411.243 & 1323.534\\

 &  & 5 - 8 & 2491.163 & 2610.437 & 2543.388 & 2519.515 & 2744.042 & 2757.741 & 2376.254 & 2617.772 & 2586.980\\

\multirow{-9}{*}{\raggedright\arraybackslash Colorado River} & \multirow{-3}{*}{\raggedright\arraybackslash Width} & 9 - 16 & 3973.402 & 4295.378 & 4104.924 & 4143.596 & 4669.426 & 5032.387 & 3960.281 & 4371.034 & 4203.610\\

\cmidrule{1-12}
 &  & 1 - 4 & 0.142 & 0.119 & 0.090 & 0.137 & 0.118 & 0.087 & 0.147 & 0.126 & 0.094\\

 &  & 5 - 8 & 0.142 & 0.127 & 0.094 & 0.154 & 0.128 & 0.093 & 0.161 & 0.140 & 0.108\\

 & \multirow{-3}{*}{\raggedright\arraybackslash RMSCE} & 9 - 16 & 0.142 & 0.125 & 0.094 & 0.156 & 0.135 & 0.104 & 0.167 & 0.151 & 0.122\\

 &  & 1 - 4 & 0.815 & 0.799 & 0.797 & 0.818 & 0.803 & 0.799 & 0.796 & 0.791 & 0.791\\

 &  & 5 - 8 & 0.813 & 0.796 & 0.793 & 0.816 & 0.801 & 0.797 & 0.782 & 0.786 & 0.787\\

 & \multirow{-3}{*}{\raggedright\arraybackslash MC} & 9 - 16 & 0.807 & 0.793 & 0.789 & 0.815 & 0.798 & 0.792 & 0.789 & 0.788 & 0.790\\

 &  & 1 - 4 & 80590.413 & 73827.493 & 68606.156 & 79251.155 & 72739.310 & 68290.714 & 73894.067 & 71544.580 & 67439.274\\

 &  & 5 - 8 & 95500.962 & 89447.215 & 85468.483 & 94526.506 & 87201.223 & 82871.135 & 83812.960 & 83612.539 & 81158.561\\

\multirow{-9}{*}{\raggedright\arraybackslash Electricity Demand} & \multirow{-3}{*}{\raggedright\arraybackslash Width} & 9 - 16 & 108864.409 & 105105.546 & 101477.482 & 103072.893 & 96090.745 & 93630.205 & 89254.761 & 87136.970 & 90483.484\\

\cmidrule{1-12}
 &  & 1 - 4 & 0.138 & 0.102 & 0.065 & 0.132 & 0.096 & 0.062 & 0.130 & 0.099 & 0.062\\

 &  & 5 - 8 & 0.148 & 0.112 & 0.072 & 0.138 & 0.107 & 0.069 & 0.134 & 0.107 & 0.068\\

 & \multirow{-3}{*}{\raggedright\arraybackslash RMSCE} & 9 - 16 & 0.148 & 0.117 & 0.077 & 0.142 & 0.113 & 0.075 & 0.138 & 0.111 & 0.072\\

 &  & 1 - 4 & 0.816 & 0.800 & 0.799 & 0.805 & 0.796 & 0.799 & 0.792 & 0.791 & 0.799\\

 &  & 5 - 8 & 0.812 & 0.796 & 0.797 & 0.806 & 0.795 & 0.799 & 0.791 & 0.789 & 0.802\\

 & \multirow{-3}{*}{\raggedright\arraybackslash MC} & 9 - 16 & 0.811 & 0.794 & 0.796 & 0.804 & 0.796 & 0.798 & 0.787 & 0.789 & 0.802\\

 &  & 1 - 4 & 14.647 & 13.906 & 13.659 & 14.642 & 14.588 & 14.584 & 14.425 & 15.296 & 15.300\\

 &  & 5 - 8 & 15.383 & 14.649 & 14.395 & 15.554 & 15.307 & 15.335 & 15.655 & 16.389 & 16.823\\

\multirow{-9}{*}{\raggedright\arraybackslash NO2 Concentration} & \multirow{-3}{*}{\raggedright\arraybackslash Width} & 9 - 16 & 15.628 & 15.415 & 15.304 & 15.915 & 15.968 & 16.218 & 15.928 & 17.375 & 17.933\\
\bottomrule
\end{tabular}}
\endgroup{}
\end{table}

\section{Experiment Implementation Detail}\label{app:implementation}
\subsection{Base Forecaster}
We evaluate three publicly released TSFM checkpoints: Chronos-Bolt (Small) (\textit{amazon/chronos-bolt-small})~\citep{chronos2}, Moirai-2.0-R (Small) (\textit{Salesforce/moirai-2.0-R-small})~\citep{moirai2}, and TimesFM-2.5 (200M) (\textit{google/timesfm-2.5-200m-pytorch})~\citep{timesfm}. For each dataset we use a rolling-origin evaluation in which the base TSFM receives a context window of $256$ consecutive observations and generates $16$ step-ahead predictive quantiles at decile levels $p \in \{0.10, 0.20, \ldots, 0.90\}$ at each forecast origin. The forecast origin advances by one step at a time through the test period.

\subsection{Calibration Protocol}
Multi-horizon rolling-origin forecasts generate a separate residual sequence for each forecast horizon $h$, with potentially different error-scale dynamics across horizons. We therefore treat each horizon as a separate calibration problem. VCPI fits a separate GARCH model for each horizon $h$ using only the $h$-step-ahead residuals, as described in Appendix~\ref{app:vcpi_algorithm}. We apply the same horizon-separating convention to the conformal baselines, running each method independently for each $h$ using only the corresponding $h$-step-ahead residuals and calibration scores. All methods share a common rolling calibration window: at forecast origin $t$, the calibration set for horizon $h$ comprises the $h$-step-ahead forecast errors with observed target times $u=t-W+1,\ldots,t$, where $W=256$. All compared methods can be thought of as post-hoc calibration methods as they consume only TSFM-generated predictions and the corresponding observed values; no method re-fits or queries the underlying TSFM.

\subsection{Native Prediction Intervals}
The three TSFMs (Chronos~\citep{chronos2}, Moirai~2~\citep{moirai2}, TimesFM~2.5~\citep{timesfm}) each output predictive quantiles at $K=9$ decile levels.
The native prediction interval at nominal level $1-\alpha$ is formed by taking the $(\alpha/2)$-th and $(1-\alpha/2)$-th predictive quantiles. Because the available quantile grid spans $[0.10,\,0.90]$, the widest constructible interval corresponds to $\alpha=0.2$, i.e., 80\% coverage, which we treat as the primary evaluation level. Results at $\alpha\in\{0.10,\,0.05\}$ are reported in Tables~\ref{tab:cov_90} and~\ref{tab:cov_95}. Indeed, native intervals are omitted at these levels because the required quantiles fall outside the decile grid.

\subsection{VCPI}
At each forecast origin $t$, the base predictive scale $\hat\sigma_{t+h}(\mathcal F_t)$ for horizon $h$ is estimated by the least-squares projection in Equation~\eqref{eq:quant_optim}, with the Gaussian reference distribution $\Phi$ (i.e.\ the standard normal), so that the reference quantiles are $z_k=\Phi^{-1}(p_k)$ for $k=1,\ldots,K$. The normalized residuals $e_u^{(h)}= \{Y_u-\hat f_u(\mathcal F_{u-h})\}/ \hat\sigma_u(\mathcal F_{u-h})$, for $u=t-W+1,\ldots,t$, are then passed to the GARCH stage.

The variance equation form and the innovation distribution are selected among three candidate specifications, standard GARCH, shown in Equation~\eqref{eq:sGARCH}, GJR-GARCH~\citep{gjrGARCH} and EGARCH~\citep{egarch}, all at GARCH order $(1,1)$. The GJR-GARCH(1,1) augments the variance recursion with an indicator for negative shocks,
\begin{align*}
    s_t^2 = \omega + (a+\lambda\,\mathbf{1}\{e_{t-1}<0\})\,e_{t-1}^2+b\,s_{t-1}^2,
\end{align*}
where $\omega>0$, $a\geq0$, $b\geq0$, $\lambda\geq0$, and $a+\lambda/2+b<1$ for covariance stationarity. $\lambda$ captures the asymmetric effect of the positive versus negative residuals. The EGARCH(1,1) parameterizes the log conditional variance to avoid positivity constraints,
\begin{align*}
    \log s_t^2
    = \omega+a\!\left(|\xi_{t-1}|-\mathbb E[|\xi_{t-1}|]\right)
    +\gamma_E\,\xi_{t-1}+b\log s_{t-1}^2,
\end{align*}
where $\xi_{t-1}=e_{t-1}/s_{t-1}$ is the standardized residual. Candidate models are evaluated on a held-out 20\% validation sub-slice of the calibration window, ranked first by coverage error and then by interval score. If all specifications fail to converge, the scale correction falls back to the sample standard deviation of the calibration residuals. We use the \texttt{rugarch} package~\citep{rugarch} in R to fit the GARCH models.

\subsection{Adaptive Conformal Inference (ACI)}
ACI is implemented in-house using R following the algorithm description in their original paper~\citep{conformal2}. The miscoverage parameter $\alpha_t$ is initialized at the nominal level $\alpha$ and updated after each observation by
\[
  \alpha_{t+1}
  =\alpha_t+\gamma\bigl(\alpha-\mathbf{1}\{Y_t\notin
  \widehat C_{t,\alpha}^{ACI}(\mathcal F_{t-1})\}\bigr),
\]
where $\gamma=0.005$ is the learning rate. This value matches the default used in \citet{conformal2}. That paper does not provide explicit guidance for choosing $\gamma$ in new settings; we therefore also report a sensitivity analysis over a range of learning rates in Table~\ref{tab:aci_sensitivity}.

\subsection{Quantile Tracking (QT)}
QT is implemented in-house using R following the algorithm description in their original paper~\citep{conformal_pid}. The quantile estimate $\hat q_t^{QT}(\mathcal F_{t-1})$ is updated after each observation by a subgradient step,
\[
  \hat q^{QT}_{t+1}(\mathcal F_t)
  =\hat q^{QT}_t(\mathcal F_{t-1})
  +\gamma\bigl(\mathbf{1}\{Y_t\notin
  \widehat C_{t,\alpha}^{QT}(\mathcal F_{t-1})\}-\alpha\bigr),
\]
where the step size is $\gamma=0.05\times\overline{|R|}$ and $\overline{|R|}$ denotes the mean absolute residual computed over the calibration window. Scaling the learning rate by the mean absolute residual follows the convention used in \citet{conformal_pid}, which does not provide explicit guidance for choosing the multiplier in new settings. We therefore also report sensitivity results in Table~\ref{tab:qt_sensitivity}.

\subsection{EnbPI}
We use the official Python implementation of EnbPI by the authors~\citep{spci}. At each forecast origin, $B=10$ bootstrap resamples of the calibration window are used to train an ensemble of linear regression base learners, and the resulting leave-one-out residuals determine the conformal quantile.

\subsection{SPCI}
We use the official Python implementation of SPCI by the authors~\citep{spci}. SPCI constructs prediction intervals by fitting a random forest to predict the conditional quantile of forecast residuals autoregressively. At each forecast origin, the model takes the $L$ most recent leave-one-out residuals as input features and outputs a predicted quantile, which is then used to form the prediction interval. The lag order of the autoregressive feature vector is set to $L=20$. The number of bootstrap ensemble members is set to $B=10$, and the conditional quantile is estimated via the empirical quantile of the random forest leaf distribution rather than through explicit quantile regression.

\subsection{Evaluation}\label{app:eval_details}
The metrics used to evaluate interval performance are detailed below. For a fixed forecast horizon $h$, let $\{(Y_t,\widehat C_{t,\alpha}(\mathcal F_{t-h}))\}_{t=1}^T$ denote the observed outcomes and associated prediction intervals, where $\widehat C_{t,\alpha}(\mathcal F_{t-h})=[\hat L_{t,h,\alpha},\hat U_{t,h,\alpha}]$ is the $(1-\alpha)$ prediction interval at time $t$ based on information available at time $t-h$.
We report three complementary measures: marginal coverage, local calibration error, and interval width. Marginal coverage is defined as
\[
\mathrm{MC}_h
=
\frac{1}{T}\sum_{t=1}^T \mathbf{1}\{Y_t\in \widehat C_{t,\alpha}(\mathcal F_{t-h})\},
\]
which measures the average empirical coverage over the full evaluation period. A method is marginally well calibrated when $\mathrm{MC}_h$ is close to the nominal level $1-\alpha$. Since marginal coverage averages over time, it does not distinguish between a method that is uniformly well calibrated and one that alternates between periods of undercoverage and overcoverage.
To assess local calibration, we compute rolling coverage over a window of length $\ell=100$,
\[
\widehat{\mathrm{MC}}_{t,h}^{(\ell)}
=
\frac{1}{\ell}\sum_{u=t-\ell+1}^t \mathbf{1}\{Y_u\in \widehat C_{u,\alpha}(\mathcal F_{u-h})\},
\]
and define the root mean squared coverage error (RMSCE) as
\[
\mathrm{RMSCE}_h
=
\sqrt{
\frac{1}{|\mathcal T_{\ell,h}|}
\sum_{t\in\mathcal T_{\ell,h}}
\left(
\widehat{\mathrm{MC}}_{t,h}^{(\ell)}-(1-\alpha)
\right)^2
},
\]
where $\mathcal T_{\ell,h}$ is the set of endpoints of valid rolling windows containing $\ell$ consecutive evaluated targets at horizon $h$. RMSCE measures how far local empirical coverage deviates from the target level over time. By squaring the deviation before averaging, it penalizes larger local calibration errors more heavily. In particular, RMSCE is useful for detecting persistent undercoverage or delayed adaptation following changes in forecast uncertainty.
To summarize efficiency, we report the median interval width,
\[
\mathrm{MW}_h
=
\operatorname{median}_{1\le t\le T}(\hat U_{t,h,\alpha}-\hat L_{t,h,\alpha}).
\]
This quantity reflects the typical size of the prediction intervals. We use the median rather than the mean because it is less sensitive to occasional extreme widths, which can occur when a method reacts aggressively to temporary bursts of volatility or isolated forecast failures.

\subsection{Compute}
All experiments were run on a shared academic SLURM cluster using CPU nodes with Intel Xeon processors; no GPU was used. Each of the three TSFM forecasting jobs (one per model, covering all four datasets under the rolling-origin protocol described above) was submitted as an independent SLURM job and allocated a single CPU with 16~GB of memory. Each model was run inside its own conda environment because the three official inference libraries  (\texttt{chronos} for Chronos-Bolt,\texttt{uni2ts} for Moirai, and \texttt{timesfm} for TimesFM) have incompatible dependencies. A single GARCH(1,1) fit completes in a few seconds, but the rolling protocol requires fitting at each forecast origin, separately for each of the 16 horizons. Aggregated over $\sim 3{,}000$ origins per dataset, 16 horizons, and three base TSFMs across four datasets, calibration alone accounts for a non-trivial share of the total compute. Combining all these factors, the total compute was on the order of 100 CPU-days.
\newpage

\end{document}